\documentclass[11pt,a4paper]{article}
\pdfoutput=1

\usepackage{jheppub}
\usepackage[T1]{fontenc}
\usepackage{extarrows}

\title{Hydrodynamics dual to Einstein-Gauss-Bonnet gravity: all-order gradient resummation}
\author{Yanyan Bu,}
\author{Michael Lublinsky,}
\author{and Amir Sharon}
\affiliation{Department of Physics, Ben-Gurion University of the Negev, \\
Beer-Sheva 84105, Israel}

\emailAdd{yybu@post.bgu.ac.il}
\emailAdd{lublinm@bgu.ac.il}
\emailAdd{sharon.amir24@gmail.com}

\abstract{Relativistic hydrodynamics dual to Einstein-Gauss-Bonnet gravity in asymptotic $\textrm{AdS}_5$ space is under study. To linear order in the amplitude of the fluid velocity and temperature, we derive the fluid's stress-energy tensor via an all-order resummation of the derivative terms. Each order is accompanied by new transport coefficients, which all together could be compactly absorbed into two functions of momenta, referred to as viscosity functions. Via inverse Fourier transform, these viscosities appear as memory functions in the constitutive relation between components of the stress-energy tensor.}


\arxivnumber{1504.01370}
\begin{document}
\maketitle

\flushbottom
\section{Introduction}\label{section1}
Fluid dynamics~\cite{fluid1,fluid2} is an effective long-wavelength description of most classical or quantum many-body systems at nonzero temperature. For neutral fluids in flat space, the hydrodynamic equations are derivable from conservation of the fluid's stress-energy tensor $T_{\mu\nu}$,
\begin{equation}\label{hydro eq}
\partial^{\mu}T_{\mu\nu}=0.
\end{equation}
For relativistic fluids, $T_{\mu\nu}$ is conveniently written as
\begin{equation}\label{stress tensor}
T_{\mu\nu}=(\varepsilon+P)u_{\mu}u_{\nu}+P\eta_{\mu\nu}+\Pi_{\langle\mu\nu\rangle},
\end{equation}
where $\varepsilon$, $u_{\mu}$ are the fluid's energy density and four-velocity field, whereas $\eta_{\mu\nu}$ stands for Minkowski metric tensor. The pressure $P$ is specified through equation of state $P=P(\varepsilon)$, calculable from underlying microscopic theory. Deviations from thermal equilibrium are collectively encoded in dissipation tensor $\Pi_{\mu\nu}$,
\begin{equation}\label{def pi}
\Pi_{\langle\mu\nu\rangle}\equiv \frac{1}{2}\mathcal{P}_{\mu}^{\alpha} \mathcal{P}_{\nu}^{\beta} \left(\Pi_{\alpha\beta}+\Pi_{\beta\alpha}\right)-\frac{1}{3} \mathcal{P}_{\mu\nu} \mathcal{P}^{\alpha\beta} \Pi_{\alpha\beta}.
\end{equation}
where $\mathcal{P}_{\mu\nu}=\eta_{\mu\nu}+u_{\mu}u_{\nu}$ is a projector on spatial directions.

At each order in derivative expansion, $\Pi_{\mu\nu}$ is fixed by thermodynamics and symmetries, up to some transport coefficients. The latter have to be calculated from microscopic description of the fluid rather than from hydrodynamics itself. In what follows, we focus on conformal fluids in 4D Minkowski spacetime, so the condition $T_{\mu}^{\mu}=0$ implies $\varepsilon=3P$. The first order derivative expansion gives the Navier-Stokes term
\begin{equation}\label{ns hydro}
\Pi_{\mu\nu}^{\textrm{NS}}=\,-2\eta_{_{0}}\partial_{\mu}u_{\nu},
\end{equation}
where $\eta_{_{0}}$ is a shear viscosity. At second order, there are five additional  transport coefficients~\cite{0712.2451,0712.2456}.

AdS/CFT correspondence~\cite{hep-th/9711200} relates strong coupling physics of gauge theories with large number of colors $N$ to weakly coupled gravity in (asymptotic) AdS space. As a particular example, it maps hydrodynamic fluctuations of a boundary fluid into long-wavelength gravitational perturbations of a stationary black brane in asymptotic AdS space~\cite{hep-th/0104066,hep-th/0205052,hep-th/0210220}. Viscosity and all other transport coefficients could be computed from the gravity side of the correspondence. The ratio of $\eta_{_{0}}$ over the entropy density $s$ was computed in~\cite{hep-th/0104066,hep-th/0205052,hep-th/0405231}
\begin{equation}\label{ratio}
\frac{\eta_{_{0}}}{s}=\frac{1}{4\pi}
\end{equation}
and was found to be universal for all gauge theories with Einstein gravity duals~\cite{hep-th/0311175,0809.3808,0808.3498}. The value~(\ref{ratio}) was further conjectured to be Nature's lower bound for $\eta_{_{0}}/s$~\cite{hep-th/0309213}.

The relativistic Navier-Stokes hydrodynamics is well known to violate causality, that is it admits propagation of signal faster than the speed of light. Inclusion of any finite number of additional derivative terms in $\Pi_{\mu\nu}$ would not render the theory into causal. All-order derivative resummation is necessary to restore causality. In~\cite{1406.7222,1409.3095,1502.08044}, we built upon the work of~\cite{0905.4069} and linearly resummed derivative terms (see~\cite{0906.4423,1103.3452,1203.0755,1302.0697,1411.1969} for boost invariant case) for fluids dual to pure Einstein gravity. In a parametrically controllable approximation, where we only collect terms linear in amplitude of the fluid velocity, $\Pi_{\mu\nu}$ has a compact form,
\begin{equation}\label{diss tensor}
\Pi_{\mu\nu}=-2\eta\left[u^{\alpha}\partial_{\alpha},\partial^{\alpha}\partial_{\alpha} \right]\partial_{\mu}u_{\nu}-\zeta \left[u^{\alpha}\partial_{\alpha}, \partial^{\alpha} \partial_{\alpha}\right]\partial_{\mu}\partial_{\nu} \partial^{\alpha} u_{\alpha}.
\end{equation}
Here $\eta$ and $\zeta$ are derivative operators, which upon expansion in a series would generate the usual gradient expansion. Thanks to linearization, we can study these operators in Fourier space, via replacement $\partial_{\mu}\longrightarrow \left(-i\omega, i \vec{q}\right)$. Then the operators $\eta$ and $\zeta$ are turned into functions of momenta and are referred to as viscosity functions. In momentum space, the constitutive relation~(\ref{diss tensor}) is
\begin{equation}\label{diss tensor ft}
\Pi_{\mu\nu}(\omega,q)=-2\eta(\omega,q^2)iq_{\mu}u_{\nu}(\omega,q)+\zeta(\omega,q^2) iq_{\mu}q_{\nu} q^{\alpha}u_{\alpha}(\omega,q).
\end{equation}
The viscosity functions $\eta$ and $\zeta$ were computed exactly in~\cite{1406.7222,1409.3095,1502.08044} and were observed to vanish at very large momenta, signaling restoration of causality in the dual CFT. For self-consistency of presentation we will flash these results in section~\ref{section3} below.

Vanishing of the viscosities at large frequencies is a necessary condition for causality restoration. To better understand the physical role of the viscosity functions, we turn them into memory functions via inverse Fourier transform of~(\ref{diss tensor ft})
\begin{equation}
\Pi_{\mu\nu}(t)=-\int_{-\infty}^{\infty}dt^{\prime}\left[2\tilde{\eta}(t-t^{\prime},q^2) \partial_{\mu}u_{\nu}(t^{\prime})+\tilde{\zeta}(t-t^{\prime},q^2)\partial_\mu\partial_{\nu} \partial^{\alpha}u_{\alpha}(t^{\prime})\right],
\end{equation}
where
\begin{equation}
\tilde{\eta}(t,q^2)\equiv \int_{-\infty}^{\infty}\frac{d\omega}{\sqrt{2\pi}} \eta(\omega,q^2) e^{-i\omega t},~~~~
\tilde{\zeta}(t,q^2)\equiv \int_{-\infty}^{\infty}\frac{d\omega}{\sqrt{2\pi}} \zeta(\omega,q^2) e^{-i\omega t}.
\end{equation}
Here $\tilde{\eta}(t,q^2)$ and $\tilde{\zeta}(t,q^2)$ are memory functions in mixed-$(t,q^2)$ representation. Causality\footnote{In the limit $N\to\infty$ and $\lambda\to\infty$, causality of $\mathcal{N}=4$ super-Yang-Mills plasma was analyzed~\cite{0805.2570} by studying pole structures of retarded correlators.} requires memory functions to have no support in the future: $\tilde{\eta}(t-t^\prime)\sim\Theta(t-t^\prime)$ and $\tilde{\zeta}(t-t^\prime)\sim \Theta(t-t^\prime)$.
In other words, the current $\Pi_{\mu\nu}(t)$ at time $t$ should be affected by the state of the system in the past only. So, for a causal theory $\Pi_{\mu\nu}$ becomes
\begin{equation}\label{mem}
\Pi_{\mu\nu}(t)=\int_{-\infty}^t dt^{\prime}\left[\tilde{\eta}(t-t^{\prime},q^2) \partial_{\mu} u_{\nu}(t^{\prime})+\tilde{\zeta}(t-t^{\prime},q^2)\partial_{\mu} \partial_{\nu} \partial^{\alpha}u_{\alpha}(t^{\prime})\right].
\end{equation}
As has been discussed in~\cite{1502.08044}, a typical memory function-based formalism~\cite{km,fluid2}, as a phenomenological model, would set the low limit of integration in~(\ref{mem}) to zero, turning thus defined hydrodynamics into a well-posed initial value problem.

\begin{figure}[htbp]
\centering
\includegraphics[scale=0.50]{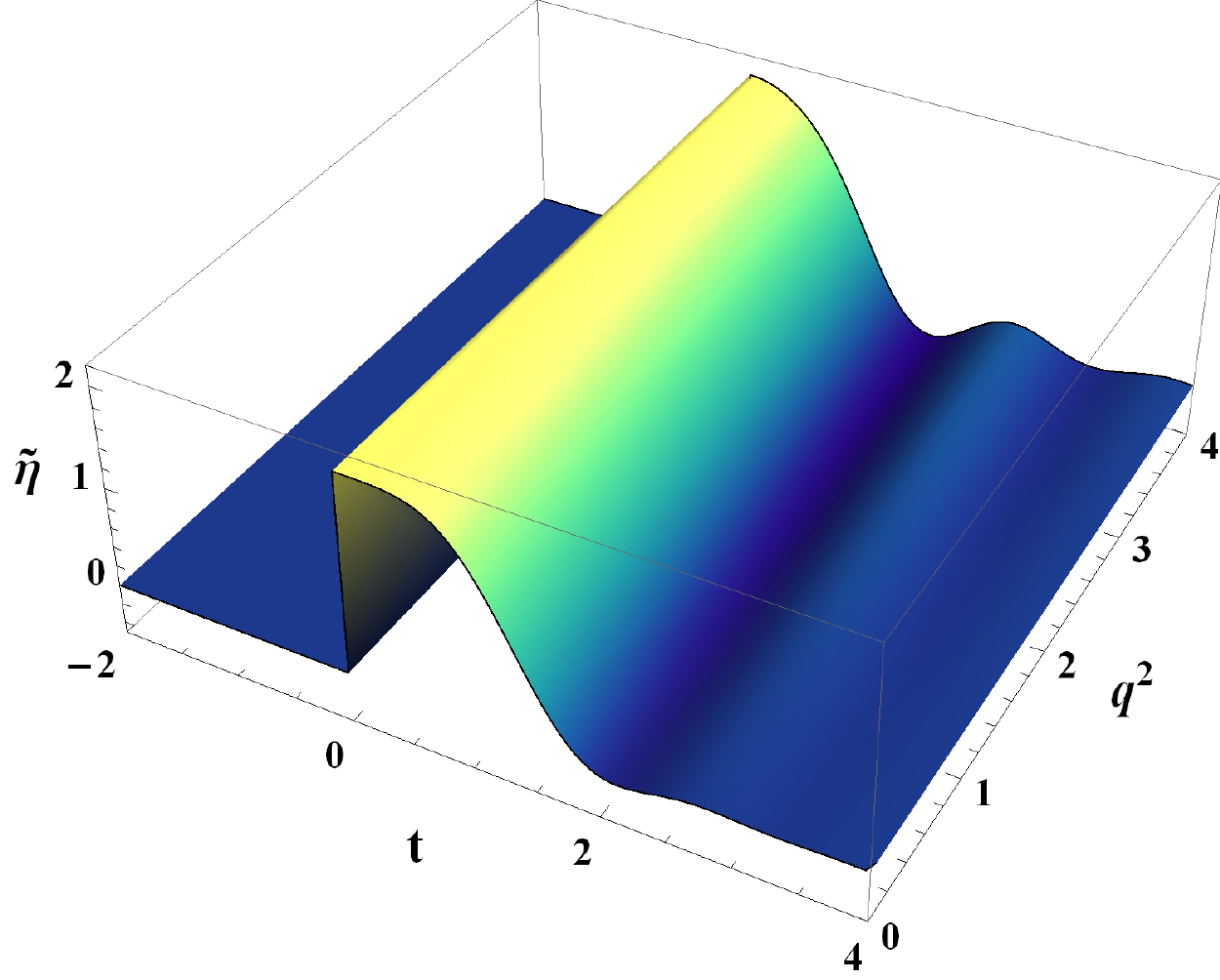}
\includegraphics[scale=0.80]{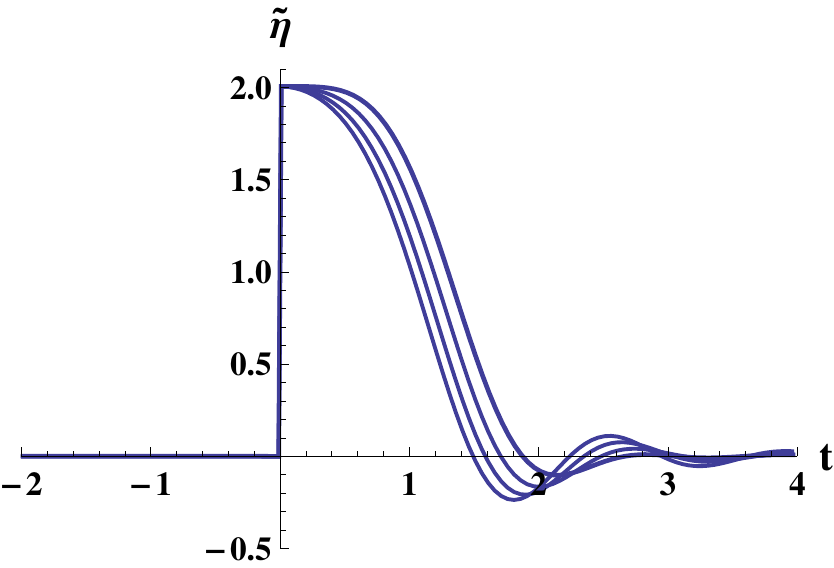}
\caption{Memory function $\tilde{\eta}(t,q^2)$ evaluated in~\cite{1502.08044} for hydrodynamics dual to pure Einstein gravity. Left: 3D plot as function of $t$ and $q^2$. Right: 2D plots as functions of time $t$: different curves display results with different $q^2$ (from the rightmost: $q^2=0,1,2,3$).}
\label{figure0}
\end{figure}

In~\cite{1502.08044}, the memory function $\tilde\eta(t)$ was evaluated from exact computations in the dual Einstein gravity. It was indeed found to be proportional (up to numerical noise) to $\Theta(t)$ as could be seen from Figure~\ref{figure0} (in units $\pi T=1$).

Beyond $N\to\infty$ and 't Hooft coupling $\lambda\to\infty$ limits, the ratio~(\ref{ratio}) gets corrected. Finite $N$ or $\lambda$ corrections arising from stringy or quantum effects introduce, beyond Einstein gravity, terms with higher derivatives of curvature. Exact forms of these terms generated in string theory are not known in general: the first higher derivative correction is expected to be the curvature squared. Of particular interest is a ghost-free Gauss-Bonnet combination, which generates equations of motion of second order only. Adding a Gauss-Bonnet term to the gravitational action is equivalent to introducing some $\mathcal{O}(1/N)$ corrections in the dual gauge theory, whereas the Gauss-Bonnet coupling $\alpha$ is related to the difference between two central charges of the dual CFT.

From the string theory point of view, the Einstein-Gauss-Bonnet (EGB) gravity should be considered as phenomenological effective low energy theory. One may, however, consider EGB gravity on its own, as a UV complete theory. Still applying the rules of the AdS/CFT correspondence, one finds that the Gauss-Bonnet correction violates the lower bound~(\ref{ratio})~\cite{0712.0743,0712.0805} (see also~\cite{0812.2521,0811.1665,0901.1421,0901.3848,0903.2834,0903.3244}). Non-perturbative Gauss-Bonnet corrections to second order transport coefficients in conformal fluids dual to EGB gravity were considered in~\cite{1112.5345,Grozdanov:2015asa,1412.5685}. Furthermore, causality of the dual CFT sets constraints on possible values of the Gauss-Bonnet coupling. In~\cite{0712.0805,0802.3318,0906.2922,0911.4257}, the coupling $\alpha$ (stripped of units) was constrained to be
\begin{equation}\label{alpha bounds}
-\frac{7}{72}\leq \alpha \leq \frac{9}{200},
\end{equation}
where the lower (upper) bound was obtained by requiring the front-velocity in the sound (scalar) channel of the dual CFT not to exceed the speed of light. Positivity of energy flux in thought experiments done in conformal colliders~\cite{0803.1467} also constrains values of $\alpha$~\cite{0907.1625,0910.5347,0911.3160}. Remarkably, constraints on $\alpha$ from causality and positivity of energy flux were found to match~\cite{0907.1625,0910.5347,0911.3160,0911.4257}. Stability of the dual plasma also sets constraints on $\alpha$~\cite{0808.2354,0903.2527}. More recently, causality violating effects due to higher derivative corrections to Einstein-Hilbert action were discovered in high energy scattering processes of gravitons
off shock waves~\cite{1407.5597} and strings off branes~\cite{1502.01254}. Pure EGB gravity was concluded to be a-causal for $\alpha$ of order one. Causality is restored by adding an infinite tower of extra massive particles with spins higher than two~\cite{1407.5597}.

In this work, we would like to explore the effects of the Gauss-Bonnet corrections on transport coefficients, beyond known results at first and second order. To this goal we consider hydrodynamics dual to EGB gravity and calculate Gauss-Bonnet correction to viscosity functions $\eta$ and $\zeta$. Given previous constraints on $\alpha$, we limit our study to small $\alpha$ only. To linear order in $\alpha$, the fluid's energy density and pressure are
\begin{equation}\label{e+p}
\varepsilon=3P= 3\left(1+3\alpha\right) \left(\pi T\right)^4.
\end{equation}
The entropy density is evaluated from $s=dP/dT$
\begin{equation}
s=4\pi(1+3\alpha)\left(\pi T\right)^3.
\end{equation}
In the hydrodynamic limit, the viscosity functions are expandable in momenta,
\begin{equation}\label{pert exp}
\begin{split}
\eta(\omega,q^2)=&\left(1-5\alpha\right)+ \frac{1}{2} \left[\left(2-\ln{2}\right)- \left(21-5\ln{2}\right)\alpha\right] i\omega-\left[\frac{1}{48}\left(6\pi-\pi^2+24 \right. \right.\\
&\left.\left.-36\ln{2}+12\ln^2{2}\right)- \underline{6.53(280)} \alpha \right]\omega^2 -\left(\frac{1}{8}-\underline{2.11(320)}\alpha\right)q^2+ \cdots,\\
\zeta(\omega,q^2)=&\frac{1}{12}\left[(5-\pi-2\ln{2}) +\left(15\pi-87 +30\ln{2} \right) \alpha\right] +\cdots,
\end{split}
\end{equation}
where the fluid's temperature is normalized to $\pi T=1$ and all the momenta are set to be measured in these units. For a positive $\alpha$, the first term in $\eta$ yields violation of the viscosity to entropy bound~\cite{0712.0743,0712.0805,0804.2453,1012.0174},
\begin{equation}
\frac{\eta_{_{0}}}{s}=\frac{1}{4\pi}\left(1-8\alpha\right).
\end{equation}
The second term in $\eta$ is the relaxation time, calculated in~\cite{1112.5345,1211.1979}. The remaining terms are new third order transport coefficients. The underlined terms are our numerical results for $\alpha$-corrected pieces. To resum the derivative terms to all orders, we numerically compute the viscosity functions for generic $\omega$ and $q^2$.
The viscosity functions are formally expanded in $\alpha$
\begin{equation}
\eta=\eta^{(0)}+\alpha\,\eta^{(1)}+\mathcal{O}\left(\alpha^2\right),~~~~\zeta=\zeta^{(0)} +\alpha\, \zeta^{(1)}+\mathcal{O}\left(\alpha^2\right),
\end{equation}
where $\eta^{(0)}$ and $\zeta^{(0)}$ are the viscosity functions computed for pure Einstein gravity in~\cite{1406.7222,1409.3095}. The results of this calculation, particularly new results on $\eta^{(1)}$ and $\zeta^{(1)}$, are presented in subsection~\ref{subsubsection322}. When Fourier transformed into memory functions, we find that the Gauss-Bonnet correction $\tilde{\eta}^{(1)}(t)$ (and also $\tilde{\zeta}^{(1)}(t)$) is also vanishing at negative times (see Figure~\ref{figure6})\footnote{In the first version of this preprint, we made a wrong statement on causality violation based on numerical  Fourier transform, which was later found to be lacking sufficient accuracy.}.


In section~\ref{section2}, we present the holographic setup. A boosted black hole solution of the EGB gravity in asymptotic AdS$_5$ space is introduced.  Following~\cite{0712.2456}, gravitational perturbation is induced by locally varying boost velocity and black hole temperature. We then parameterize additional bulk metric corrections in terms of ten functions $h$, $k$, $j_i$ and $\alpha_{ij}$, which are both functions of holographic coordinate and functionals of the fluid velocity $u_{\mu}$. The boundary stress-energy tensor is read off from holographic renormalization, being expressed in terms of near-boundary behavior of $h$, $k$, $j_i$ and $\alpha_{ij}$. In section~\ref{section3}, we solve the Einstein equations for the metric corrections. Thanks to linearization in the velocity amplitude, all bulk metric corrections can be  decomposed in the basis formed from $u_i$. As a result, in Fourier space, the Einstein equations turn into second order ordinary differential equations for decomposition coefficients. Solutions to these equations reveal the information about the viscosities. We then discuss effects of the Gauss-Bonnet correction on the viscosity functions. Section~\ref{section4} is devoted to summary and discussion. Some computational details are provided in Appendix~\ref{appendix}.

\section{Holographic setup for Einstein-Gauss-Bonnet gravity}\label{section2}
Our representation is largely based on~\cite{0801.1021}. We start from the EGB gravity with a negative cosmological constant $\Lambda=-6/l^2$ in 5D spacetime manifold $\mathcal{M}$,
\begin{equation}\label{bulk action}
S=\frac{1}{16\pi G_{N}}\int_{\mathcal{M}} d^5x\sqrt{-g}\left(R-2\Lambda+\alpha l^2 L_{\text{GB}} \right)+S_{\text{sur}}+S_{\text{c.t.}},
\end{equation}
where the Gauss-Bonnet term $L_{\text{GB}}$ is
\begin{equation}
L_{\text{GB}}=R_{MNPQ}R^{MNPQ}-4R_{MN}R^{MN}+R^2.
\end{equation}
We use a mostly plus signature for the bulk metric $g_{MN}$. To have a well-defined variational principle, the surface term $S_{\text{sur}}$ computed in~\cite{GH1,hep-th/0208205} was added to~(\ref{bulk action}),
\begin{equation}
S_{\text{sur}}=\frac{1}{8\pi G_{N}}\int_{\partial\mathcal{M}}d^4x\sqrt{-\gamma}\left[K- 2\alpha l^2 \left(J+2G_{\mu\nu} K^{\mu\nu}\right)\right],
\end{equation}
where the first term is the Gibbons-Hawking surface action. The tensor $J_{\mu\nu}$ is defined as
\begin{equation}
J_{\mu\nu}=-\frac{1}{3}\left(2KK_{\mu\rho}K^{\rho}_{\nu}+K_{\rho\sigma}K^{\rho\sigma} K_{\mu\nu}-2K_{\mu\rho}K^{\rho\sigma}K_{\sigma\nu}-K^2K_{\mu\nu}\right),
\end{equation}
where $K_{\mu\nu}=\gamma_{\mu}^{\alpha}\bar{\nabla}_{\alpha}n_{\nu}$ and $\gamma_{\mu\nu}$, $n_{\mu}$ are the induced metric/outing normal vector on/to a constant $r$-slice $\partial \mathcal{M}$. The Einstein tensor $G_{\mu\nu}$ and $\bar{\nabla}_{\mu}$ are compatible with $\gamma_{\mu\nu}$.

In asymptotic AdS space, UV divergences near conformal boundary can be removed by holographic renormalization~\cite{hep-th/0112119,hep-th/0209067}. For the EGB gravity, the counter-term action $S_{\text{c.t.}}$ was first constructed in~\cite{0801.1021} following previous studies~\cite{hep-th/9806087,hep-th/9902121,hep-th/0010138},
\begin{equation}
S_{\text{c.t.}}=\frac{1}{8\pi G_{N}}\int_{\partial \mathcal{M}}d^4x \sqrt{-\gamma} \left(\delta_1- \frac{\delta_2}{2} \mathcal{R}[\gamma]\right),
\end{equation}
with the coefficients $\delta_1$ and $\delta_2$  having the forms
\begin{equation}
\begin{split}
\delta_1&=\frac{-1-8\alpha+\sqrt{1-8\alpha}}{\sqrt{4\alpha l^2}\sqrt{1- \sqrt{1-8\alpha}}} \stackrel{\alpha\to 0}{\longrightarrow} -\frac{3}{l} +\frac{\alpha}{l}+\mathcal{O}(\alpha^2),\\
\delta_2&=\frac{\sqrt{4\alpha l^2}\left(3-8\alpha-3\sqrt{1-8\alpha} \right)} {2\left(1-\sqrt{1-8\alpha}\right)^{3/2}}\stackrel{\alpha\to 0}{\longrightarrow} \frac{l}{2}+\frac{3l}{2} \alpha+\mathcal{O}(\alpha^2).
\end{split}
\end{equation}
Up to a conformal factor, the stress-energy tensor of the boundary CFT is obtained by varying~(\ref{bulk action}) with respect to $\gamma_{\mu\nu}$. The boundary stress-energy tensor is~\cite{0801.1021},
\begin{equation}
\begin{split}
T_{\mu\nu}&=\lim_{r\to\infty}\tilde{T}_{\mu\nu}(r)\\
&=-\lim_{r\to\infty}\frac{r^2}{8\pi G_{N}}\left\{K_{\mu\nu}- K\gamma_{\mu\nu}- \delta_1\gamma_{\mu\nu}-\delta_2G_{\mu\nu}-2\alpha l^2 \left(Q_{\mu\nu}-\frac{1}{3}Q \gamma_{\mu\nu} \right)\right\}.
\end{split}
\end{equation}
The tensor $Q_{\mu\nu}$ is defined as
\begin{equation}
Q_{\mu\nu}=3J_{\mu\nu}-2K\mathcal{R}_{\mu\nu}-\mathcal{R}K_{\mu\nu}+2K^{\rho\sigma} \mathcal{R}_{\rho\mu\sigma\nu}+4\mathcal{R}_{\mu\lambda}K^{\lambda}_{\nu},
\end{equation}
where the calligraphic tensor $\mathcal{R}_{\mu\rho\nu\sigma}$ is the  Riemann curvature of $\gamma_{\mu\nu}$.
For convenience, we set the overall scale of the stress tensor to one, $l=16\pi G_{N}=1$.

The field equations for the metric $g_{MN}$ are
\begin{equation}\label{field eq}
\begin{split}
0=E_{MN}\equiv\;&R_{MN}-\frac{1}{2}g_{MN}R-6g_{MN}-\frac{1}{2}\alpha g_{MN}L_{\text{GB}} \\
&+2\alpha\left(R_{MABC}R_{N}^{~~ABC}-2R_{MANB}R^{AB}-2R_{MA}R^{A}_{N}+RR_{MN}\right).
\end{split}
\end{equation}
A black hole solution with a flat boundary was found in~\cite{hep-th/0109133} following previous work~\cite{Boulware:1985wk}. In the ingoing Eddington-Finkelstein coordinate, the metric is
\begin{equation}\label{gs metric}
d\frak{s}^{2}=2N_{\#}dvdr-N_{\#}^{2}r^{2}f({\bf{b}}r)dv^{2}+r^{2}\delta_{ij}dx^{i} dx^{j},~~~i,j=1,2,3.
\end{equation}
To linear order in $\alpha$, we have
\begin{equation}
\begin{split}
N_{\#}&=1-\alpha+\mathcal{O}(\alpha^2),\\
f(r)&=1-\frac{1}{r^4}+2\alpha\left(1+\frac{1}{r^8}\right)+\mathcal{O}(\alpha^2).
\end{split}
\end{equation}
Thermodynamics of the EGB black holes was analyzed in~\cite{hep-th/0109133,Boulware:1985wk}.
The horizon radius $r_{\textrm{H}}$ and Hawking temperature $T$ are
\begin{equation}\label{horizon}
r_{H}=\frac{1-\alpha}{\bf{b}},~~~~T=\frac{1-2\alpha}{\pi {\bf{b}}}.
\end{equation}
The conformal boundary is at $r=\infty$.

To construct fluid dynamics from gravity, we follow~\cite{0712.2456}. First, the static black hole geometry~(\ref{gs metric}) is boosted along boundary directions $x^{\alpha}$ with a constant boost parameter $u_{\mu}$. Then, $u_{\mu}$ and $\bf{b}$ are promoted into arbitrary local functions of $x^{\alpha}$, resulting in an inhomogeneous geometry
\begin{equation}\label{boosted bh}
d\mathfrak{s}^{2}=-2N_{\#}u_{\mu}(x^{\alpha})dx^{\mu}dr-N_{\#}^{2}r^{2}f \left({\bf{b}}(x^{\alpha})r\right) u_{\mu}(x^{\alpha})u_{\nu}(x^{\alpha})dx^{\mu} dx^{\nu}+r^{2} \mathcal{P}_{\mu\nu} dx^{\mu} dx^{\nu},
\end{equation}
where $u_{\mu}(x^{\alpha})$ is identified with the fluid velocity and is normalized as $\eta^{\mu\nu}u_{\mu}(x^{\alpha})u_{\nu}(x^{\alpha})=-1$. In general, the metric~(\ref{boosted bh}) no longer solves the field equations~(\ref{field eq}). Suitable metric corrections to~(\ref{boosted bh}) are needed to make~(\ref{field eq}) satisfied. These corrections are dual to parts of $\Pi_{\langle\mu\nu\rangle}$.  Instead of the order-by-order boundary expansion~\cite{0712.2456}, we will collect the derivatives in a unified way, as proposed in~\cite{1406.7222,1409.3095,1502.08044} to resum all order linear structures in $T_{\mu\nu}$. We linearize $u_{\mu}(x^{\alpha})$ and $\mathbf{b}(x^{\alpha})$
\begin{equation}
u_{\mu}(x^{\alpha})=\left(-1,\epsilon\, u_{i}(x^{\alpha})\right),~~~~ {\bf{b}}(x^{\alpha})={\bf{b}}_{0}+\epsilon\,{\bf{b}}_{1}(x^{\alpha}),
\end{equation}
where $\epsilon$ is an order-counting parameter to be set to unity at the end. Subsequent calculations are accurate up to linear order in $\epsilon$. The constant ${\bf{b}}_{0}$ corresponds to equilibrium temperature. For convenience of calculation we set ${\bf{b}}_{0}=1$. This is equivalent to setting $\pi T=1-2\alpha$, whereas eventually we would like to present our results in units of $\pi T=1$. This is easily achieved by rescaling all the momenta by the corresponding $(1-2\alpha)$-factors.

The linearized version of~(\ref{boosted bh}) is
\begin{equation}\label{seed metric}
\begin{split}
ds_{\textrm{seed}}^{2}=&2N_{\#}dvdr-N_{\#}^2 r^2 f(r) dv^2 +r^{2}\delta_{ij} dx^i dx^j\\
-&\epsilon\left\{2N_{\#} u_idx^i dr +4 N_{\#}^2\left(1-\frac{4\alpha}{r^4} \right) \frac{{\bf{b}}_1}{r^2}dv^2+ 2r^{2} \left[1-N_{\#}^2 f(r)\right] u_i dvdx^i\right\}
\end{split}
\end{equation}
which is referred to as a seed metric. Formally, we write the full metric as
\begin{equation}
ds^2=g_{MN}dx^Mdx^N=ds_{\textrm{seed}}^2+ds_{\textrm{corr}}^2,
\end{equation}
where $ds_{\textrm{corr}}^2$ represents metric corrections. We choose a ``background field'' gauge~\cite{0712.2456}
\begin{equation}\label{gauge}
g_{rr}=0,~~~g_{r\mu}\propto u_{\mu},~~~\textrm{Tr}\left[\left(g^{(0)}\right)^{-1} g^{(1)}\right] =0,
\end{equation}
where $g^{(0)}$ corresponds to the first line in~(\ref{seed metric}) and $g^{(1)}$ denotes metric corrections. Under~(\ref{gauge}), $ds_{\textrm{corr}}^2$ can be parameterized in the form
\begin{equation}\label{metric corr}
ds_{\textrm{corr}}^2=
\epsilon \left\{\frac{k}{r^2}dv^2 - 3N_{\#} h dvdr + r^2h d{\vec x}^2
+ 2r^2\left[1-f(r)\right]j_i dx^i dv + r^2\alpha_{ij}dx^i dx^j \right\},
\end{equation}
where $\alpha_{ij}$ is a traceless symmetric tensor of rank two. The functions $h$, $k$, $j_i$ and $\alpha_{ij}$ depend on the holographic coordinate $r$ and, through the field equations~(\ref{field eq}), are functionals of the fluid velocity $u_{\mu}$.

Boundary conditions for the metric corrections were discussed in details in~\cite{1409.3095}. The first one is that all the metric components in~(\ref{metric corr}) are required to be regular over the whole range of $r$. Second, since the boundary metric is fixed to be $\eta_{\mu\nu}$, near $r=\infty$ we demand
\begin{equation}\label{AdS constraint}
h<\mathcal{O}\left(r^0\right),~k<\mathcal{O}\left(r^4\right),~ j_i<\mathcal{O}\left(r^4\right), ~\alpha_{ij}<\mathcal{O}\left(r^0\right).
\end{equation}
Finally, the fluid velocity $u_{\mu}$ is defined in Landau frame
\begin{equation}\label{Landau frame}
u^{\mu}T_{\mu\nu}=-\varepsilon u_{\nu} \Longrightarrow u^{\mu}\Pi_{\langle\mu\nu\rangle}=0.
\end{equation}
Under these boundary conditions, expressions for $\tilde{T}_{\mu\nu}$ greatly simplify.
We summarize them in Appendix~\ref{appendix}.

\section{From gravity to fluid dynamics}\label{section3}
In this section, we derive the stress-energy tensor of the boundary fluid by solving the field equations~(\ref{field eq}). There are fourteen independent components, which are split into ten dynamical equations and four constraints. As in~\cite{1406.7222,1409.3095,1502.08044}, our strategy will be to first solve the dynamical equations, without imposing the constraints. This turns out to be sufficient to uniquely fix the transport coefficients, or in other words we construct an ``off-shell'' stress-energy tensor of the dual fluid. The remaining four constraints are the conservation law of the stress-energy tensor. This equivalence is demonstrated in Appendix~\ref{appendix}.

\subsection{Deriving the fluid dynamics}\label{subsection31}
The dynamical equation $E_{rr}=0$ yields
\begin{equation}\label{h eq}
\left(1-4\alpha+4\alpha r^{-4}\right)\left(5\partial_rh + r\partial_r^2h\right)=0.
\end{equation}
The asymptotic  constraint $h<\mathcal{O}\left(r^0\right)$ and Landau frame convention $\Pi_{\langle00\rangle}=0$ lead to $h=0$. The dynamical equation for $k$ is read off from $E_{rv}=0$,
\begin{equation}\label{k eq}
\begin{split}
0=&\;3r^2\partial_rk-6r^4\partial u-r^3\partial_v\partial u+2\partial j+ r\partial_r \partial j + r^3\partial_i\partial_j\alpha_{ij} \\
&+\frac{\alpha}{r^4}\left[-48rk+3\left(4r^2-3r^6\right)\partial_rk+8\left(3r^8+r^4\right) \partial u +\left(5r^7-4r^3\right)\partial_v\partial u \right.\\
&~~~~~~~~\left.-4\left(3r^4+5\right)\partial j-2\left(3r^5-r\right)\partial_r\partial j-\left(5r^7 +4r^3\right) \partial_i \partial_j\alpha_{ij}\right],
\end{split}
\end{equation}
which will be solved by direct integration, once solutions for $j_i$ and $\alpha_{ij}$ are obtained.

From $E_{ri}=0$, we arrive at the dynamical equation for $j_{i}$,
\begin{equation}\label{j eq}
\begin{split}
0=&\;r\partial_r^2 j_i- 3\partial_r j_i+ r^3\partial_r\partial_j\alpha_{ij}+ r\partial^2 u_i-r\partial_i\partial u+3r^2\partial_v u_i \\
&-\frac{\alpha}{r^4}\left[\left(5r^5-2r\right)\partial_r^2 j_i - 3\left(5r^4-2\right) \partial_r j_i + 4\left(r^7+r^3\right) \partial_r \partial_j\alpha_{ij}\right.\\
&\left.+\left(5r^5+r\right)\left(\partial^2u_i-\partial_i\partial u\right) + 4\left(3r^6 + r^2\right)\partial_v u_i\right],
\end{split}
\end{equation}
which is coupled with $\alpha_{ij}$ only. For the tensor mode $\alpha_{ij}$, we find it more convenient to consider the combination $E_{ij}-\frac{1}{3}\delta_{ij}E_{kk}=0$,
\begin{equation}\label{alpha eq}
\begin{split}
0=&\;\left(r^7-r^3\right)\partial_r^2\alpha_{ij}+\left(5r^6-r^2\right)\partial_r \alpha_{ij} +2r^5 \partial_v \partial_r \alpha_{ij} + 3r^4 \partial_v \alpha_{ij} + r^3[[\alpha]]_{ij}\\
&+\left(1-r\partial_r\right)[[j]]_{ij}+\left(6r^4+2r^3\partial_v\right)\sigma_{ij} -\frac{\alpha}{r^4}\left[2\left(r^{11}-3r^3\right)\partial_r^2\alpha_{ij}\right.\\
&\left. +2\left(5r^{10}+9r^2\right)\partial_r\alpha_{ij} +2\left(3r^9+4r^5\right) \partial_v \partial_r\alpha_{ij}+\left(9r^8-4r^4\right)\partial_v\alpha_{ij}\right.\\
&\left. +4\left(r^7-3r^3\right)[[\alpha]]_{ij}+5\left(r^4+6\right)[[j]]_{ij}- \left(5r^5 + 6r\right)[[j]]_{ij} \right.\\
&\left. +\left(18r^8-8r^4\right)\sigma_{ij}+8\left(r^7+r^3\right) \partial_v \sigma_{ij} \right],
\end{split}
\end{equation}
where the notations $[[\alpha]]_{ij}$, $[[j]]_{ij}$ and $\sigma_{ij}$ are defined as
\begin{equation}\label{notations funs}
\begin{split}
&[[\alpha]]_{ij}\equiv \partial^2\alpha_{ij}-\left(\partial_i \partial_k \alpha_{jk} + \partial_j \partial_k \alpha_{ik}- \frac{2}{3}\delta_{ij} \partial_k \partial_l \alpha_{kl}\right),\\
&[[j]]_{ij}\equiv \partial_i j_j+\partial_j j_i-\frac{2}{3}\delta_{ij}\partial j,~~~
2\sigma_{ij}\equiv \partial_i u_j+\partial_j u_i-\frac{2}{3}\delta_{ij}\partial u.
\end{split}
\end{equation}

Notice that, as in~\cite{1409.3095}, source terms in~(\ref{j eq},\ref{alpha eq}) are only constructed from $u_i$. To solve these partial differential equations, we first decompose $j_i$ and $\alpha_{ij}$ in a basis formed from $u_i$,
\begin{equation}\label{decomposition}
\left\{
\begin{aligned}
j_{i}&=a\left(\partial_v,\partial^2,r\right)u_i+b\left(\partial_v,\partial^2,r\right) \partial_i\partial u,\\
\alpha_{ij}&=2c\left(\partial_v,\partial^2,r\right)\sigma_{ij}+d\left(\partial_v, \partial^2,r\right)\pi_{ij},
\end{aligned}
\right.
\end{equation}
where $\sigma_{ij}$ is defined in~(\ref{notations funs}) and $\pi_{ij}\equiv \partial_i\partial_j\partial u-\frac{1}{3}\delta_{ij} \partial^2 \partial u$. Then, in Fourier space, dynamical equations~(\ref{j eq},\ref{alpha eq}) translate into a system of second order ordinary differential equations for the decomposition coefficients
\begin{equation}\label{abcd momentum}
\left\{
\begin{aligned}
0=&\;r\partial_r^2a-3\partial_r a-\bar{q}^2 r^3\partial_rc-\bar{q}^2r-3i\bar{\omega}r^2 -\frac{\alpha}{r^4}\left[\left(5r^4-2\right)\left(r\partial_r^2a-3\partial_r a\right) \right.\\
&\left.-4\bar{q}^2\left(r^7+r^3\right)\partial_rc- \bar{q}^2\left(5r^5+r\right) -4i\bar{\omega}\left(3r^6 +r^2\right)\right],\\
0=&\; r\partial_r^2b-3\partial_rb-\frac{2}{3}\bar{q}^2r^3\partial_rd+ \frac{1}{3} r^3 \partial_rc-r-\frac{\alpha}{r^4}\left[\left(5r^4- 2\right)\left(r\partial_r^2 b -3\partial_r b \right)\right.\\
&\left.- \frac{4}{3} \left(r^7+r^3\right) \partial_r\left(2\bar{q}^2d-c\right) -\left(5r^5+r\right)\right],\\
0= &\; \left(r^7-r^3\right)\partial_r^2c+\left(5r^6-r^2\right) \partial_r c -2i\bar{\omega} r^5 \partial_r c -r\partial_ra+a-3i\bar{\omega} r^4c\\
&-i\bar{\omega} r^3+3r^4-\frac{\alpha}{r^4}\left[2\left(r^{11}-3r^3\right) \partial_r^2 c +2\left(5r^{10}+9r^2\right)\partial_rc \right.\\
&\left.-2i\bar{\omega} \left(3r^9+4r^5\right) \partial_r c -i\bar{\omega} \left(9r^8-4r^4\right)c+5\left(r^4+6\right)a \right.\\
&\left.-\left(5r^5+6r\right) \partial_r a +(9r^8-4r^4)- 4i\bar{\omega} \left(r^7+ r^3\right)\right],\\
0= &\; \left(r^7-r^3\right)\partial_r^2d+\left(5r^6-r^2\right) \partial_rd -2i\bar{\omega} r^5\partial_r d -\frac{1}{3}r^3\left(2c-q^2d\right)+2b \\
&-2r\partial_rb -3i\bar{\omega} r^4d-\frac{\alpha}{r^4}\left[2\left(r^{11}-3r^3 \right) \partial_r^2d +2\left(5r^{10}+9r^2\right)\partial_rd\right.\\
&\left.-2i\bar{\omega} \left(3r^9+ 4r^5 \right)\partial_rd - i\bar{\omega} \left(9r^8 -4r^4\right)d  -\frac{4}{3}\left(r^7-3r^3\right) \left(2c-\bar{q}^2d\right) \right.\\ &\left.+10\left(r^4+ 6 \right) b -2\left(5r^5+6r\right)\partial_rb \right].
\end{aligned}
\right.
\end{equation}
Equation~(\ref{k eq}) becomes
\begin{equation}\label{k momentum}
\begin{aligned}
0=&\;\left(1-3\alpha+\frac{4\alpha}{r^4}\right)\partial_rk-\frac{16\alpha}{r^5}k-  \left\{2r^2 -\frac{1}{3}i\bar{\omega} r-\frac{2}{3r^2}\left(a-\bar{q}^2b\right)\right.\\
&\left.- \frac{1}{3r} \left(\partial_r a-\bar{q}^2\partial_r b\right) -\frac{2}{9}\bar{q}^2 r\left(\bar{q}^2 d-2c\right)+\frac{\alpha}{r^4} \left[\frac{1}{3}i \bar{\omega} (5r^{5}-4r) \right.\right.\\
&\left.\left. +\frac{2}{3r}\left(3r^4-1\right)\left(\partial_r a-\bar{q}^2 \partial_r b\right) +\frac{2}{9}\bar{q}^2 r\left(5r^5+4r\right)\left(\bar{q}^2 d-2c\right)\right.\right.\\
&\left.\left.+\frac{4}{3r^2}\left(3r^4+5\right) \left(a-\bar{q}^2b\right)-\frac{8}{3} \left(3r^6+r^2\right)\right]\right\}\partial u.
\end{aligned}
\end{equation}
The barred momenta are defined as $\bar{\omega}\equiv (1-2\alpha)\omega$ and $\bar{q}\equiv (1-2\alpha)q$, which emerge as a result of the above
mentioned rescaling of units.

We first study the large $r$ behavior of the metric corrections, which propagates into the expression for the fluid's stress tensor. The velocity dependence of $T_{\mu\nu}$ enters via the decomposition~(\ref{decomposition}). Examining equations~(\ref{abcd momentum}) near the conformal boundary $r=\infty$, it is straightforward to show that
\begin{equation}\label{asym abcd}
\begin{split}
a&\xlongrightarrow{r\to\infty}-i\bar{\omega}\left(1+\alpha\right)r^3+\mathcal{O} \left(\frac{1}{r}\right),~~~~~~~~~~~~b\xlongrightarrow{r\to\infty}- \frac{1}{3}r^2+ \mathcal{O}\left(\frac{1}{r}\right),\\
c&\xlongrightarrow{r\to\infty}\frac{1-\alpha}{r}+\frac{C_b^4\left(\bar{\omega}, \bar{q}^2\right)}{r^4}+\mathcal{O} \left(\frac{1}{r^5}\right),~~~~ d\xlongrightarrow{r\to\infty}\frac{D_b^4\left(\bar{\omega},\bar{q}^2\right)}{r^4}+ \mathcal{O}\left(\frac{1}{r^5}\right),
\end{split}
\end{equation}
where $C_b^4$ and $D_b^4$ are unknown coefficients, which cannot be determined from the asymptotic analysis alone. To compute them, we have to integrate~(\ref{abcd momentum}) over the entire bulk. Regularity of the metric components in~(\ref{metric corr}) imposes two boundary conditions at $r=r_{\textrm{H}}$, which are  sufficient to fix $C_b^4$ and $D_b^4$ uniquely. The large $r$ behavior of $k$ is
\begin{equation}\label{asym k}
k\xlongrightarrow{r\to\infty}\left\{\frac{2}{3}\left(1-\alpha\right) r^3 +
\frac{2}{3} \left(1-2\alpha\right) i\bar{\omega} r^2\right\}\partial u +\mathcal{O} \left(\frac{1}{r}\right).
\end{equation}
Boundary conditions~(\ref{AdS constraint},\ref{Landau frame}) were imposed in deriving~(\ref{asym abcd},\ref{asym k}).

Plugging~(\ref{asym abcd},\ref{asym k}) into~(\ref{stress tld1},~\ref{stress tld2},~\ref{stress tld3}), we obtain the boundary stress-energy tensor
\begin{equation}\label{emt}
\left\{
\begin{aligned}
T_{00}=&\; 3\left(1-5\alpha\right)\left(1-4{\bf{b}}_1\right),\\
T_{0i}=&\;T_{i0}=-4\left(1-5\alpha\right)u_i,\\
T_{ij}=&\;\delta_{ij}\left(1-5\alpha\right)\left(1-4{\bf{b}}_1\right)\\
&+ 4\left(1-3\alpha \right) \left[2C_b^4\left(\bar{\omega},\bar{q}^2\right) \left(1+6\alpha\right)\sigma_{ij}+D_b^4 \left(\bar{\omega},\bar{q}^2\right) \left(1+2\alpha\right)\pi_{ij}\right].
\end{aligned}
\right.
\end{equation}
Covariantization of~(\ref{emt}) gives standard expressions~(\ref{stress tensor},\ref{diss tensor}) of $T_{\mu\nu}$, with $\varepsilon$ and $P$ given by~(\ref{e+p}). The viscosity functions $\eta$ and $\zeta$ re-expressed in units of $\pi T=1$ are
\begin{equation}\label{vis formula}
\begin{split}
\eta\left(\omega,q^2\right)&=-4\left(1-3\alpha\right)C_b^4\left[\left(1-2\alpha\right) \omega,\left(1-4\alpha\right)q^2\right] \left(1+6\alpha\right), \\ \zeta\left(\omega,q^2\right)&= -4\left(1-3\alpha\right)D_b^4\left[\left(1-2\alpha\right) \omega,\left(1-4\alpha\right) q^2\right]\left(1+2\alpha\right).
\end{split}
\end{equation}

\subsection{Gauss-Bonnet corrections to the viscosity functions}\label{section32}
To determine the viscosity functions, we are now ready to fully solve the dynamical equations~(\ref{abcd momentum}). In the next subsection, we start with the hydrodynamic limit and solve~(\ref{abcd momentum}) perturbatively in momenta. In this way, we reproduce some known results in the literature and also obtain a set of new third order transport coefficients. In the subsection to follow, we address our main goal of resumming all-order derivative terms. This will be achieved by numerically solving~(\ref{abcd momentum}) for generic values of $\bar{\omega}$ and $\bar{q}^2$.

\subsubsection{Analytical results: hydrodynamic expansion}\label{subsubsection321}
We introduce power counting parameter $\lambda$ by $\bar{\omega}\rightarrow \lambda\bar{\omega}$ and $\bar{q}_i \rightarrow \lambda \bar{q}_i$, and expand the decomposition coefficients~(\ref{decomposition}) in powers of $\lambda$,
\begin{equation}\label{abcd expansion}
\begin{split}
a\left(\bar{\omega},\bar{q}_i,r\right)&=\sum_{n=0}^{\infty}\lambda^n a_n\left(\bar{\omega},\bar{q}_i,r\right),~~~ b\left(\bar{\omega},\bar{q}_i,r\right)=\sum_{n=0}^{\infty}\lambda^n b_n\left(\bar{\omega},\bar{q}_i,r\right),\\
c\left(\bar{\omega},\bar{q}_i,r\right)&=\sum_{n=0}^{\infty}\lambda^n c_n\left(\bar{\omega},\bar{q}_i,r\right),~~~ d\left(\bar{\omega},\bar{q}_i,r\right)=\sum_{n=0}^{\infty}\lambda^n d_n\left(\bar{\omega},\bar{q}_i,r\right).
\end{split}
\end{equation}
At each order in $\lambda$, there is a system of ordinary differential equations for $a_n$ etc, whose solutions are double integrals. In Appendix~\ref{appendix}, we summarize these results. Then, $C_b^4$ and $D_b^4$ are expanded as
\begin{eqnarray}\label{cd exp}
\begin{split}
C_b^4(\bar{\omega},\bar{q}^2)=&-\frac{1}{4}\left(1-8\alpha\right)-\frac{1}{8}i\bar{\omega} \left[\left(2-\ln{2}\right) -\left(23-6\ln{2}\right)\alpha\right]+ \bar{q}^2 \left[\frac{1}{32}-\underline{0.497(227)}\alpha\right]\\
&+\bar{\omega}^2\left\{\frac{1}{192}\left(6\pi-\pi^2+24-36\ln{2}+12\ln^2{2}\right)- \underline{1.56(140)}\alpha\right\}+\cdots,\\
D_b^4(\bar{\omega},\bar{q}^2)=&\frac{1}{48}\left(\pi-5+2\ln{2}\right)+\frac{1}{24} \left(41-7\pi-14\ln{2}\right) \alpha+\cdots,
\end{split}
\end{eqnarray}
where in $C_b^4$ we have only numerical results for the  linear in $\alpha$ second order terms.
The viscosities~(\ref{pert exp}) are obtained by substituting~(\ref{cd exp}) in~(\ref{vis formula}).

Taking plane wave ansatz for $u_i$ and ${\bf{b}}_1$, the conservation law $\partial^{\mu}T_{\mu\nu}=0$ results in dispersion equations
\begin{equation}\label{dispersion}
\begin{split}
&\textrm{shear wave}:~~\left(1+3\alpha\right)\omega+\frac{1}{4}iq^2\eta\left(\omega,q^2 \right)=0,\\
&\textrm{sound wave}:~\left(1+3\alpha\right)\left(q^2-3\omega^2\right)-i\omega q^2 \eta\left(\omega,q^2\right)+\frac{1}{2}i \omega q^4\zeta\left(\omega,q^2\right)=0.
\end{split}
\end{equation}
In the hydrodynamic limit, the dispersion equations~(\ref{dispersion}) could be solved perturbatively. For the lowest modes they read
\begin{eqnarray}\label{qnm}
\begin{split}
\textrm{shear wave}:~\omega=&-\frac{i}{4}\left(1-8\alpha\right)q^2-\frac{i}{32} \left[1-\log{2}+\left(8\#_1-40+16\log{2}\right)\alpha\right]q^4+\cdots,\\
\textrm{sound wave}:~\omega=&\pm\frac{q}{\sqrt{3}}-\frac{i}{6}\left(1-8\alpha\right)q^2 \pm\frac{1}{24\sqrt{3}}\left[3-2\log{2}+\left(16\log{2}-38\right)\alpha\right]q^3\\
&-\frac{i}{864}\left[\pi^2-24+24\log{2}-12\log^2{2}+\left(\#_1+144\#_2-294\right. \right.\\
&\left.\left.~~~~~~~~~-90\pi-3\pi^2+60\log{2}+36\log^2{2}\right)\alpha\right]q^4+\cdots,
\end{split}
\end{eqnarray}
where ${\#}_1=6.53(280)$ and $\#_2=2.11(320)$ are known  numerically only.
These hydrodynamic modes should agree with the lowest quasi-normal modes of the EGB gravity.

\subsubsection{Numerical results: all-order resummed hydrodynamics} \label{subsubsection322}
For generic values of $\omega$ and $q^2$, we resort to a shooting technique and solve~(\ref{abcd momentum}) numerically. Our numerical procedure is essentially the same as that of~\cite{1409.3095}. We start with a guess solution at the horizon $r=r_H$ and integrate~(\ref{abcd momentum}) until the conformal boundary $r=\infty$. Then, we fine-tune the initial guess until thus generated solution satisfies the boundary conditions at $r=\infty$.
\begin{figure}[htbp]
\centering
\includegraphics[scale=0.55]{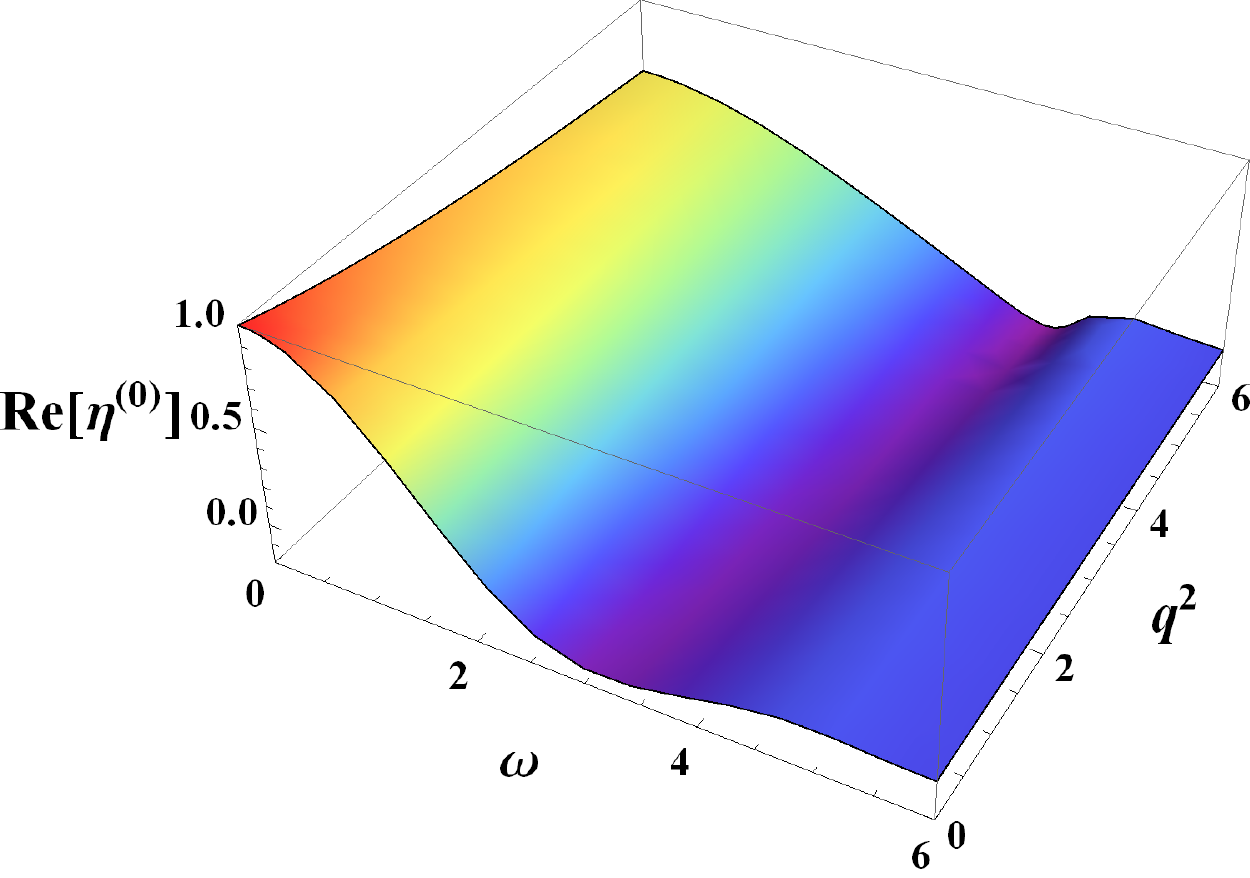}
\includegraphics[scale=0.55]{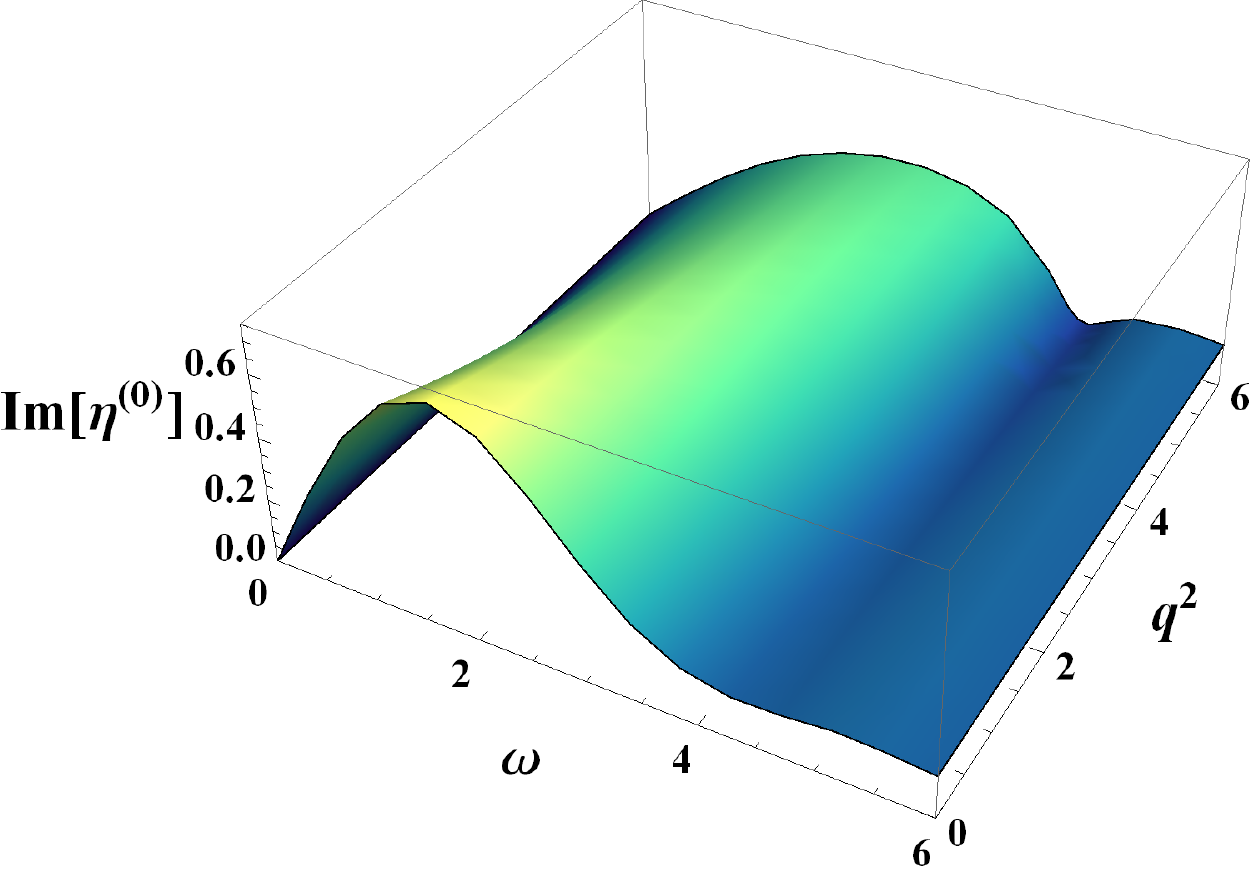}
\includegraphics[scale=0.55]{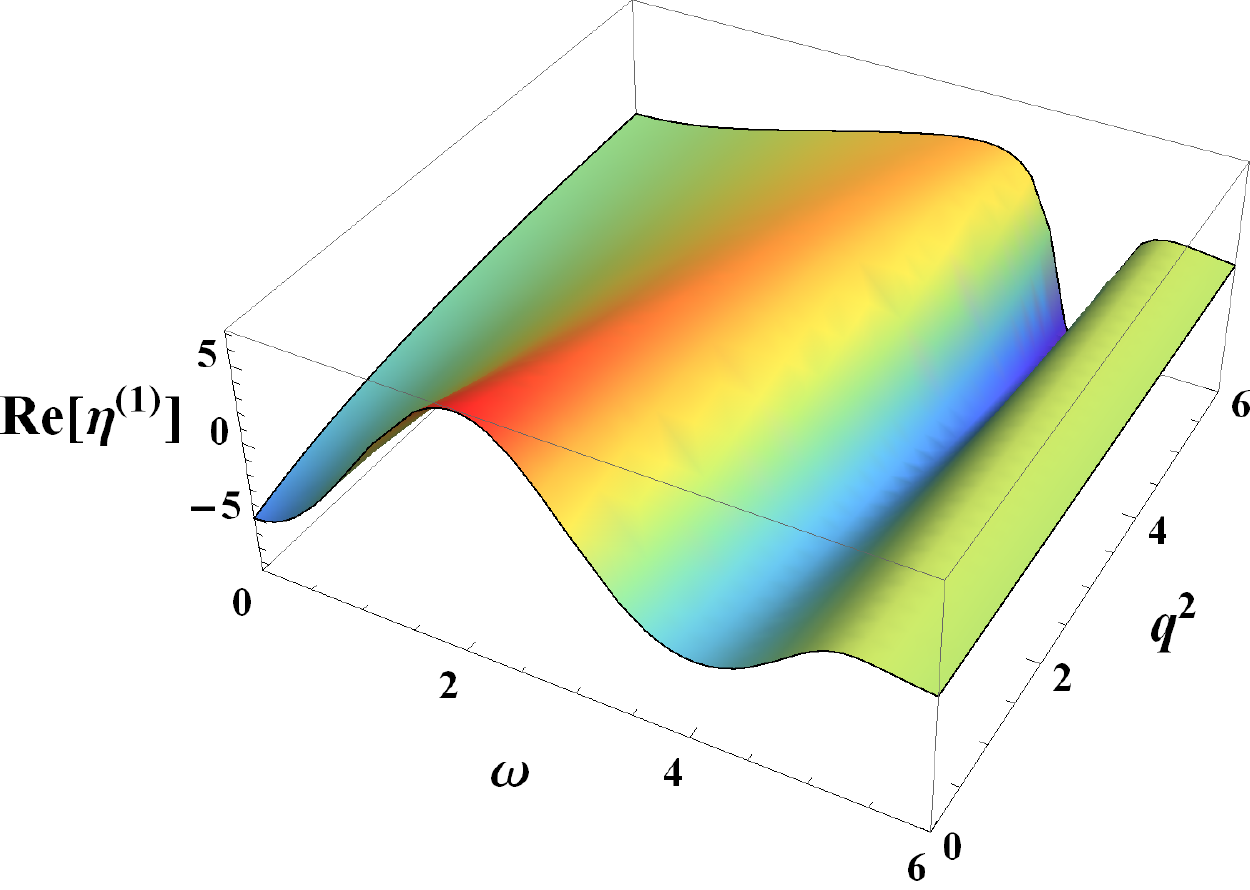}
\includegraphics[scale=0.55]{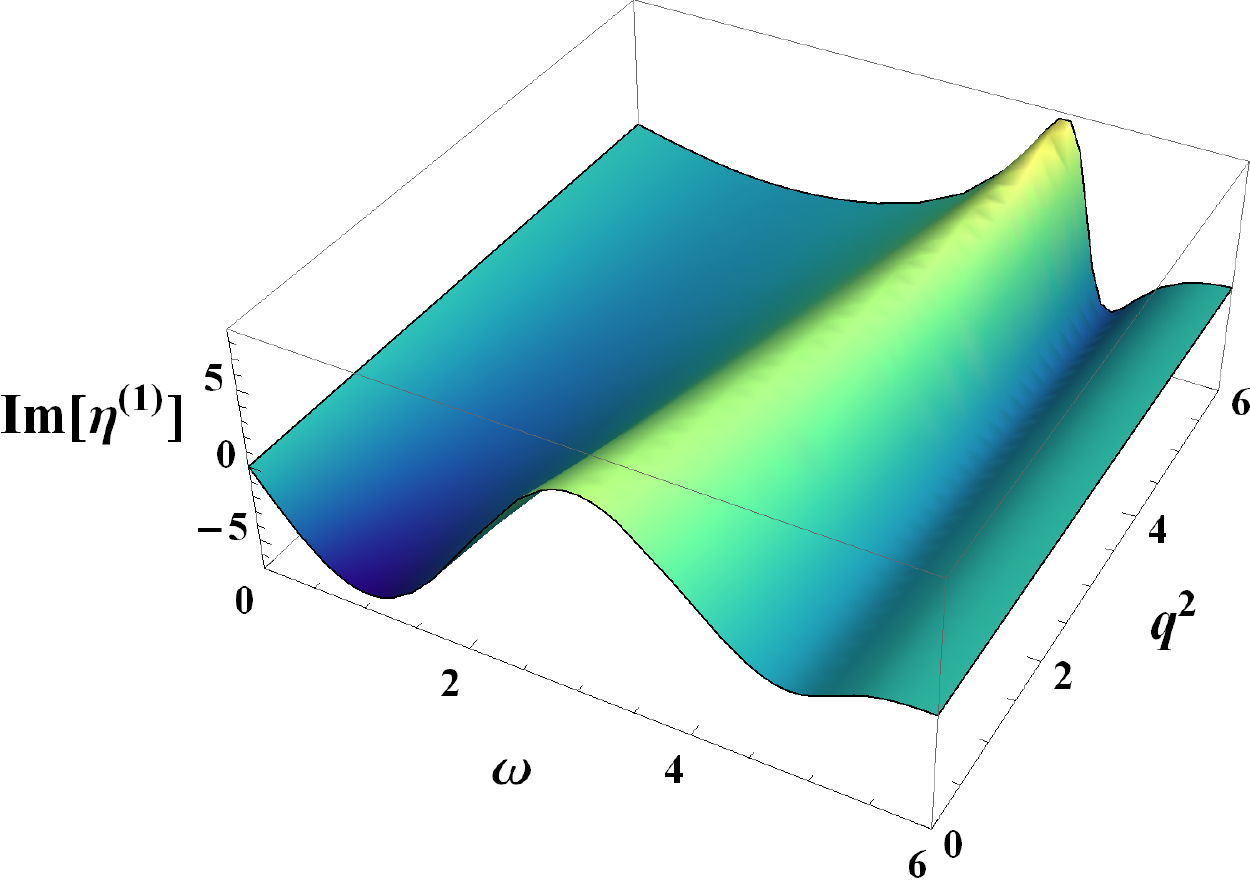}
\caption{The viscosity $\eta$ as function of $\omega$ and $q^2$.}
\label{figure1}
\end{figure}
\begin{figure}[htbp]
\centering
\includegraphics[scale=0.55]{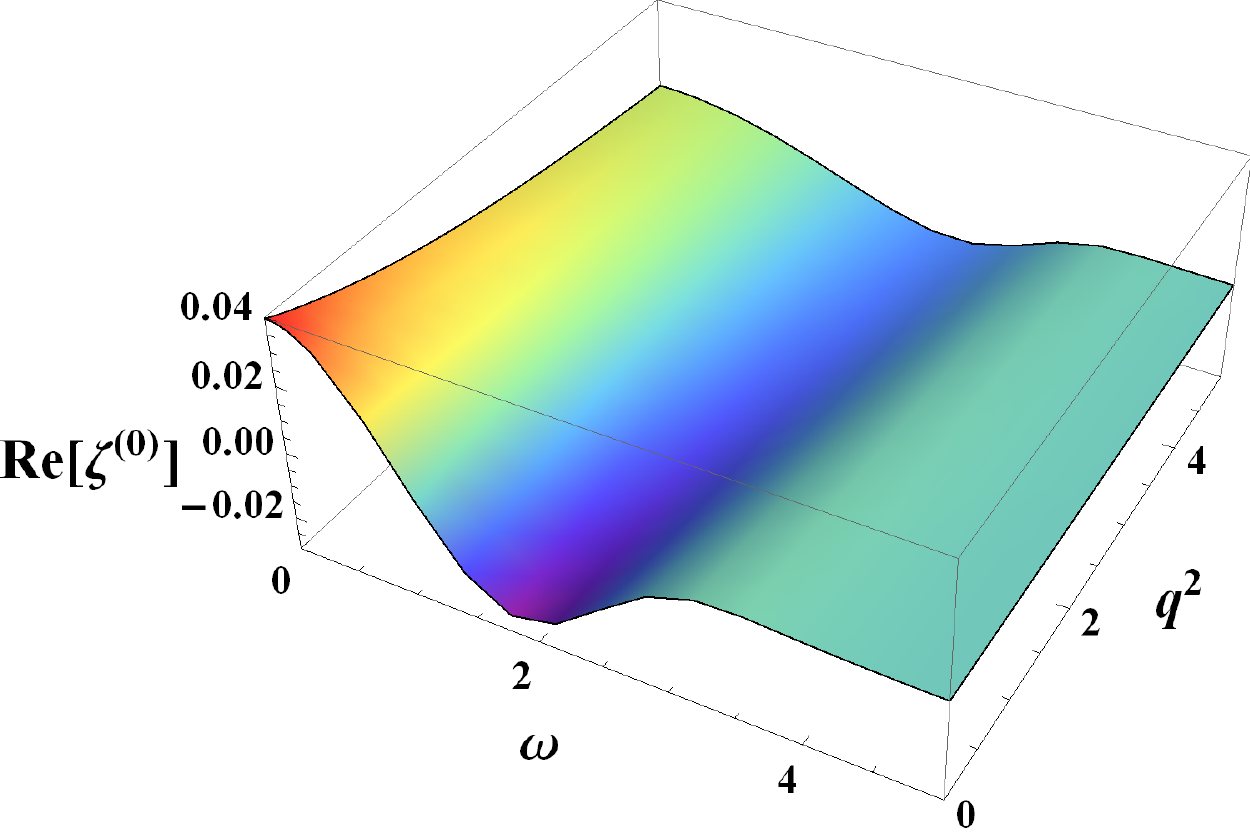}
\includegraphics[scale=0.55]{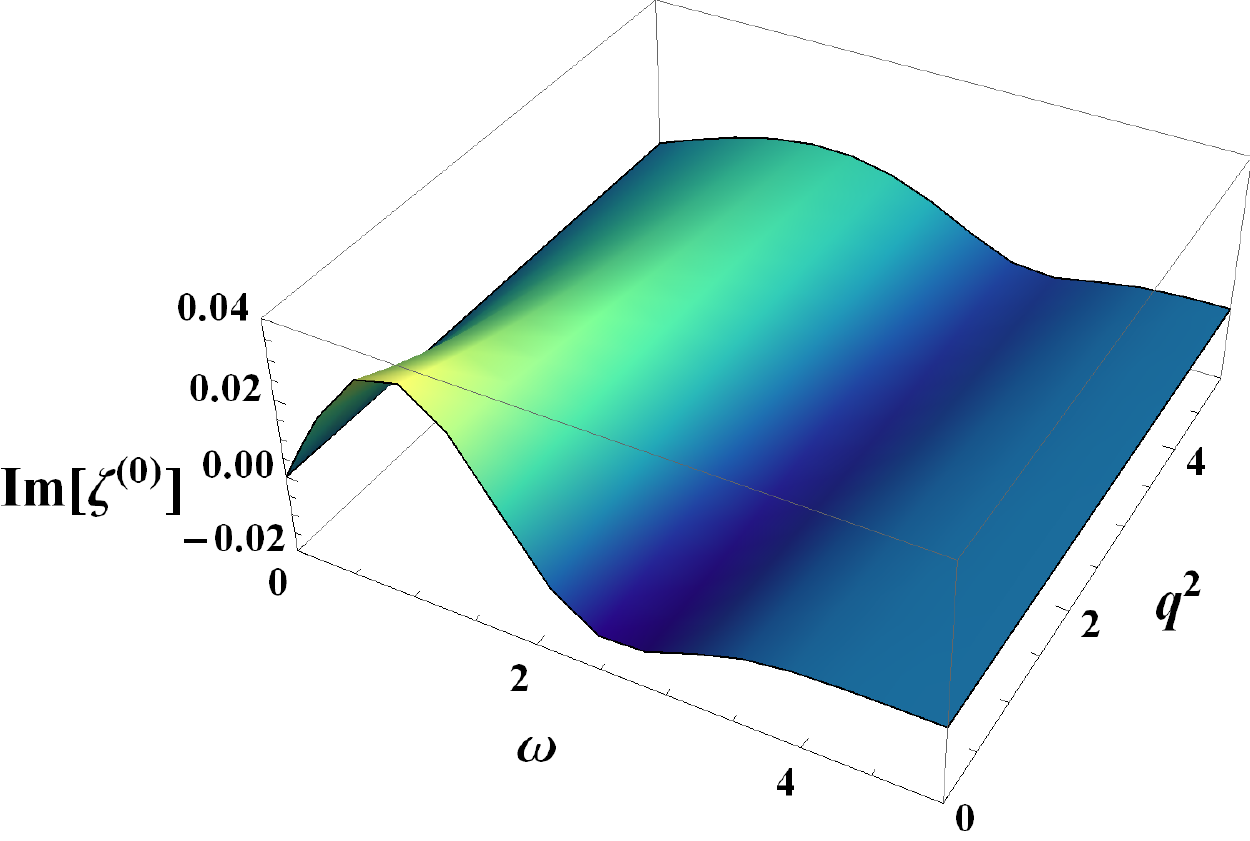}
\includegraphics[scale=0.55]{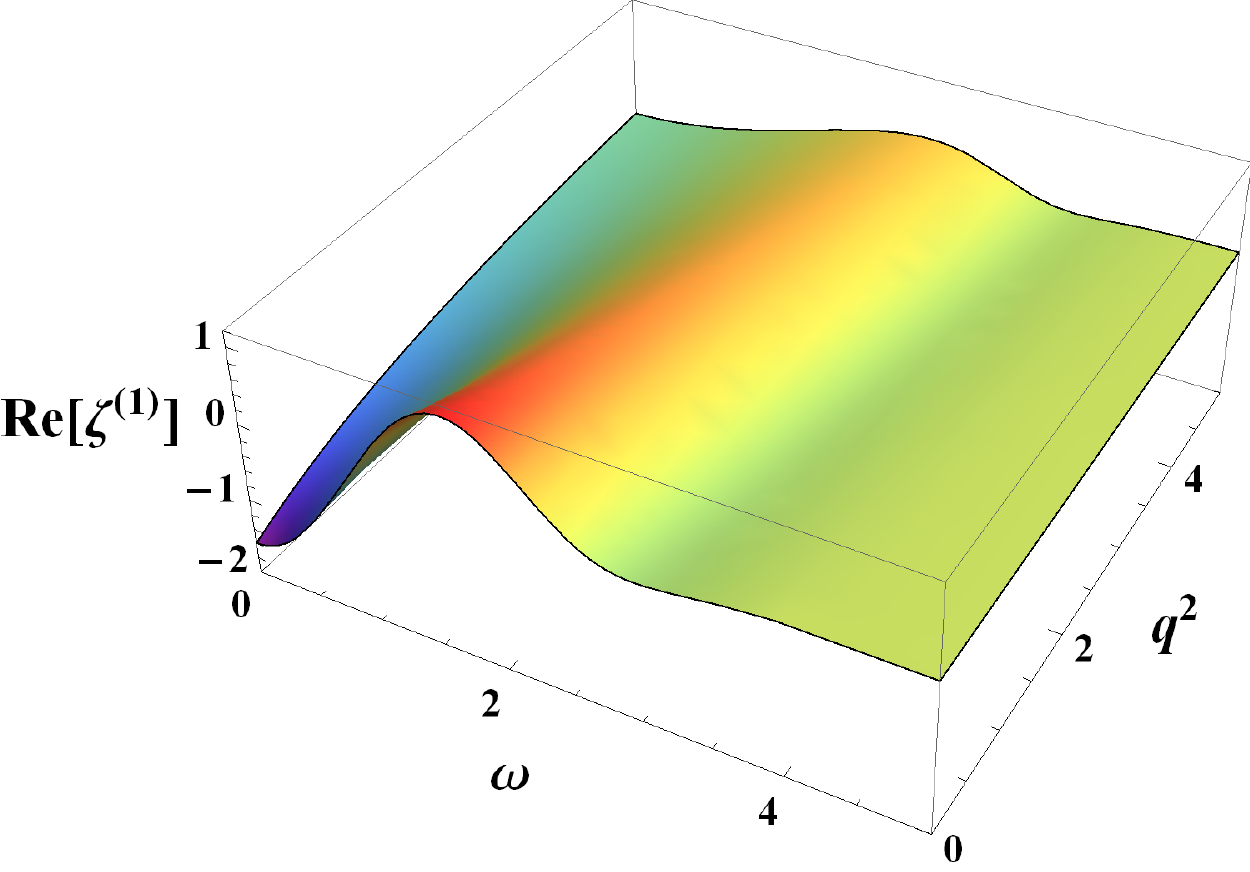}
\includegraphics[scale=0.55]{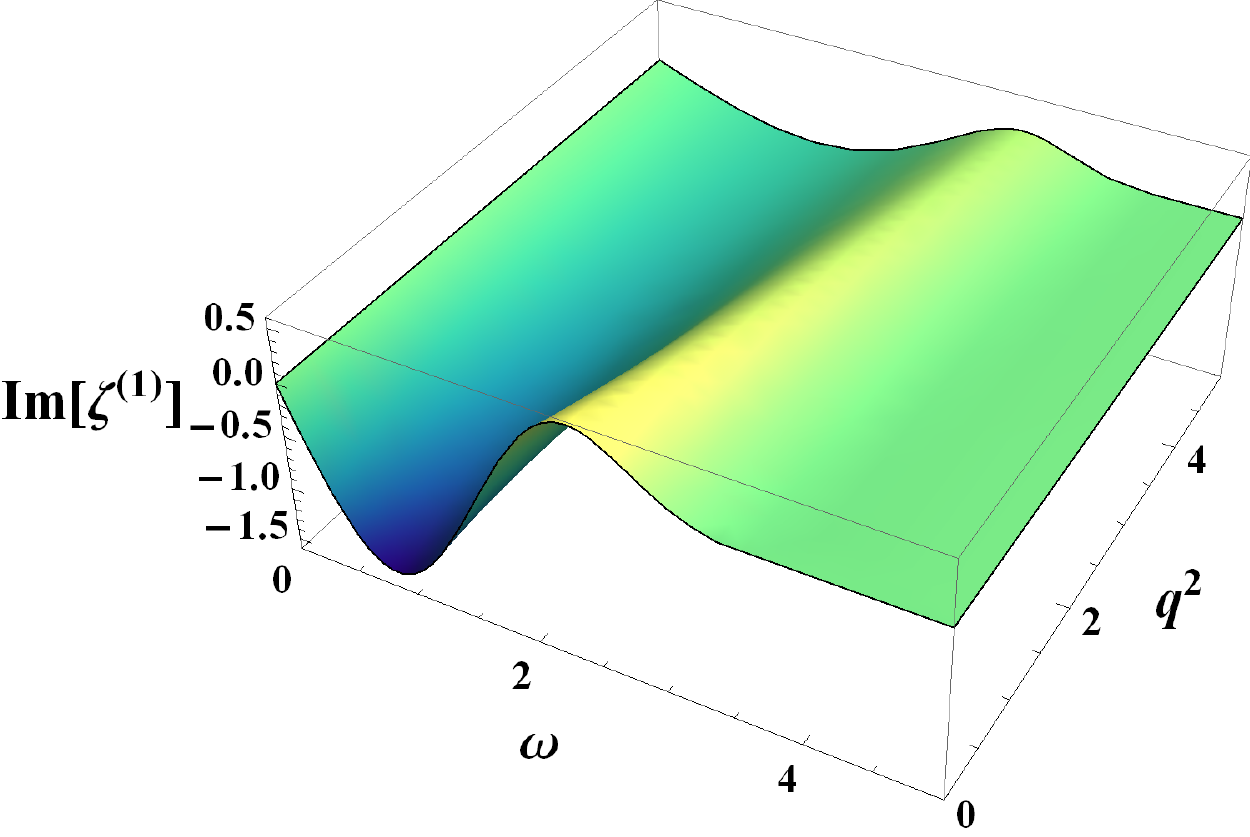}
\caption{The viscosity $\zeta$ as function of $\omega$ and $q^2$.}
\label{figure2}
\end{figure}
\begin{figure}[htbp]
\centering
\includegraphics[scale=0.80]{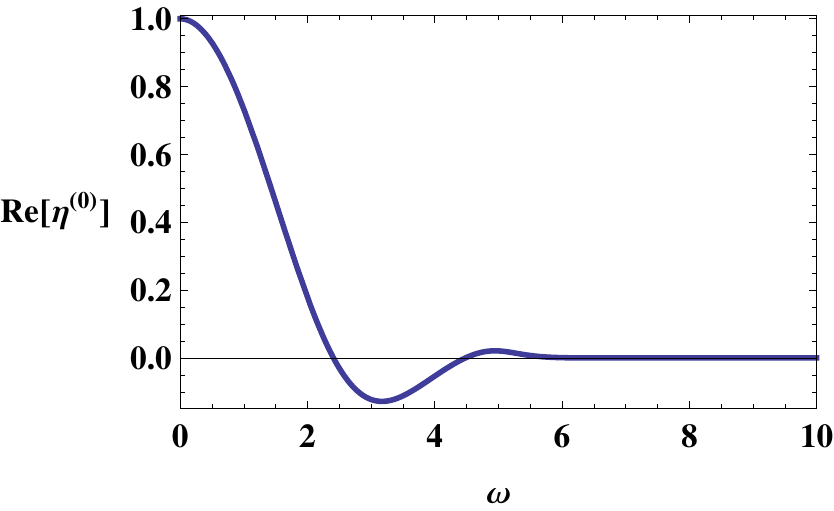}
\includegraphics[scale=0.80]{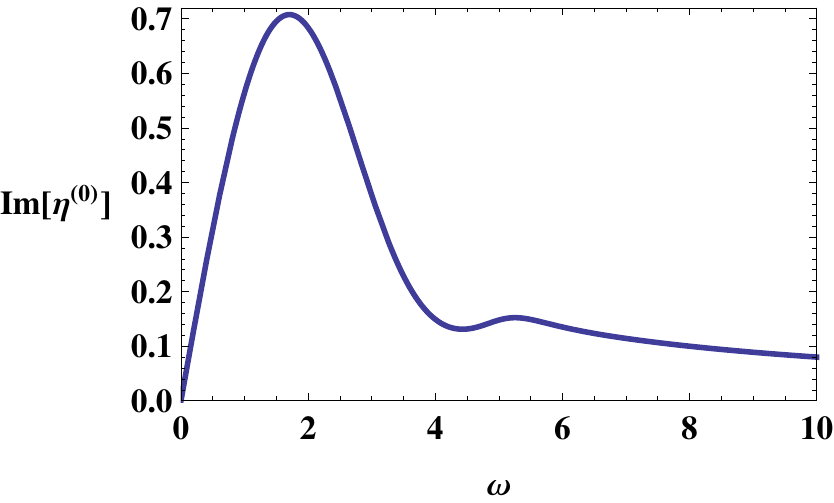}
\includegraphics[scale=0.80]{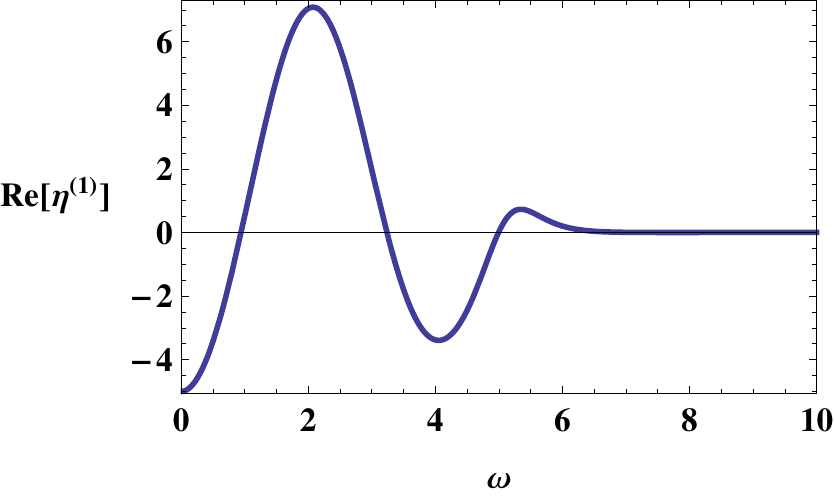}
\includegraphics[scale=0.80]{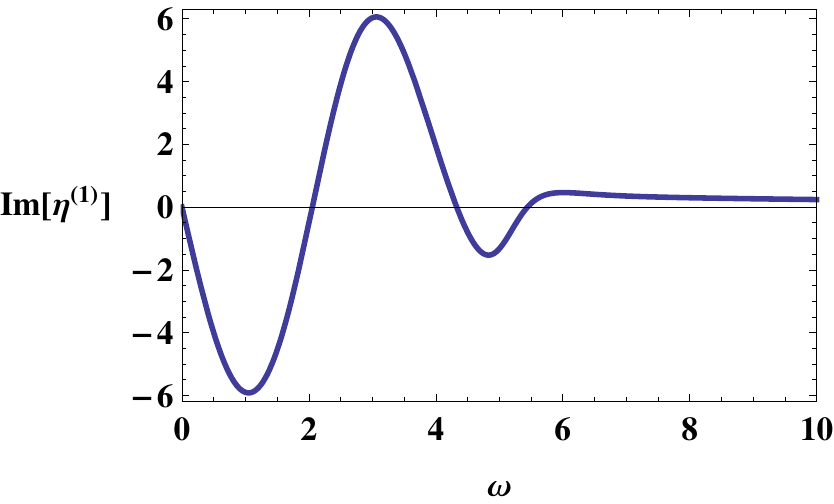}
\caption{The viscosity $\eta$ as function of $\omega$ with $q=0$.}
\label{figure3}
\end{figure}
\begin{figure}[htbp]
\centering
\includegraphics[scale=0.83]{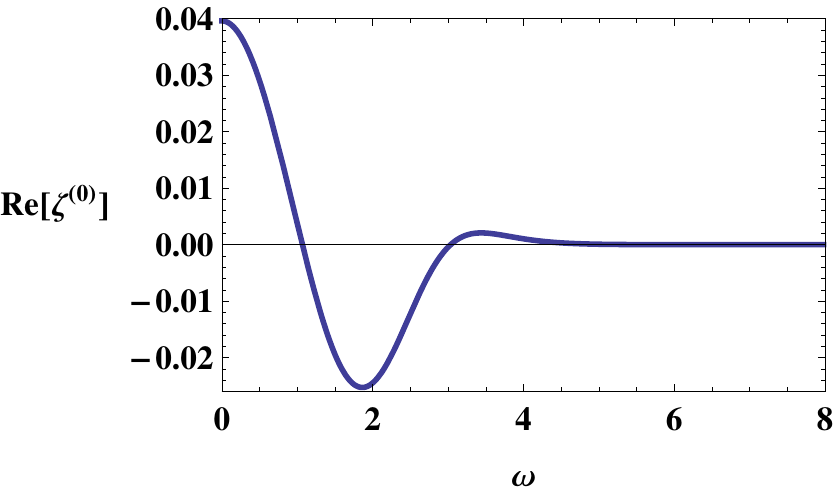}
\includegraphics[scale=0.80]{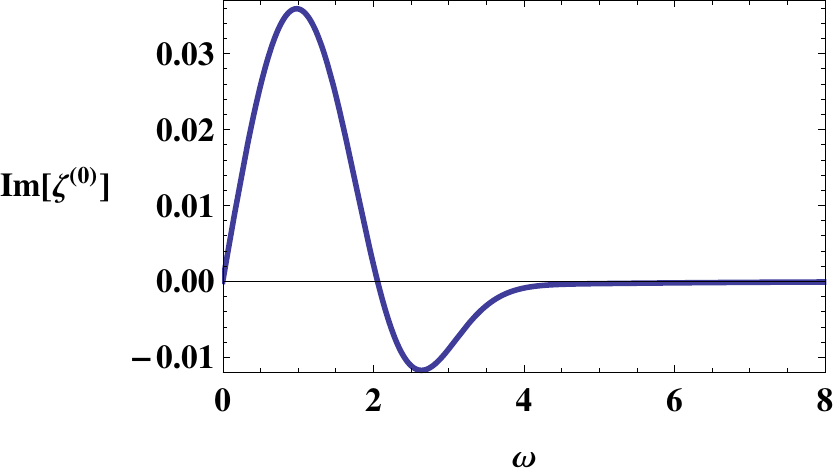}
\includegraphics[scale=0.80]{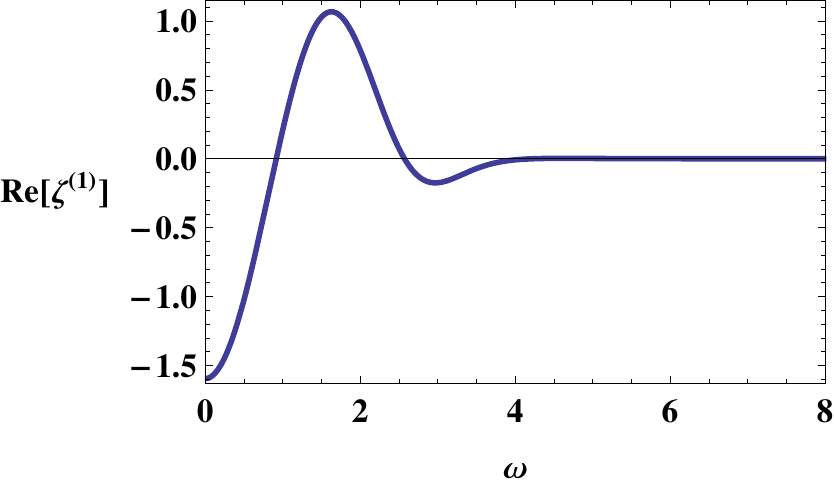}
\includegraphics[scale=0.80]{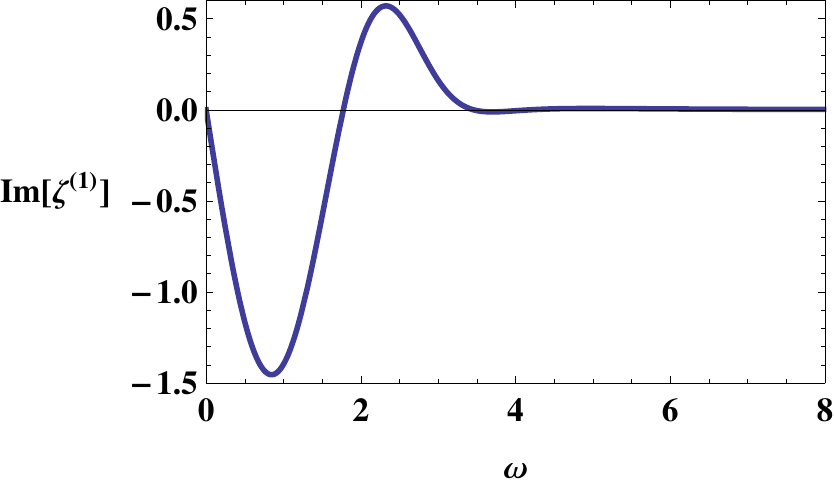}
\caption{The viscosity $\zeta$ as function of $\omega$ with $q=0$.}
\label{figure4}
\end{figure}
\begin{figure}[htbp]
\centering
\includegraphics[scale=0.80]{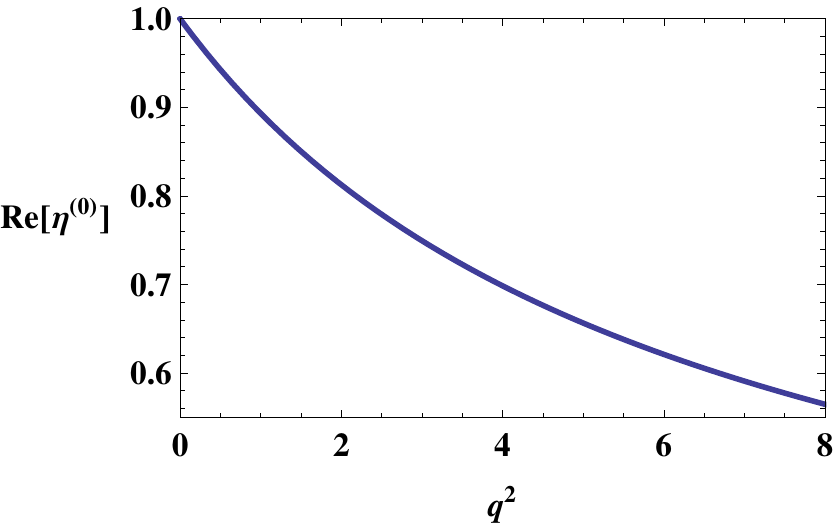}
\includegraphics[scale=0.83]{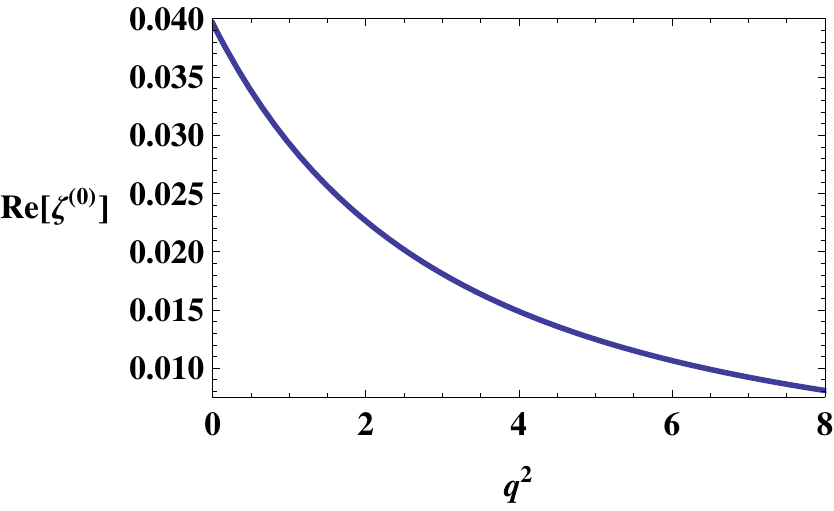}
\includegraphics[scale=0.80]{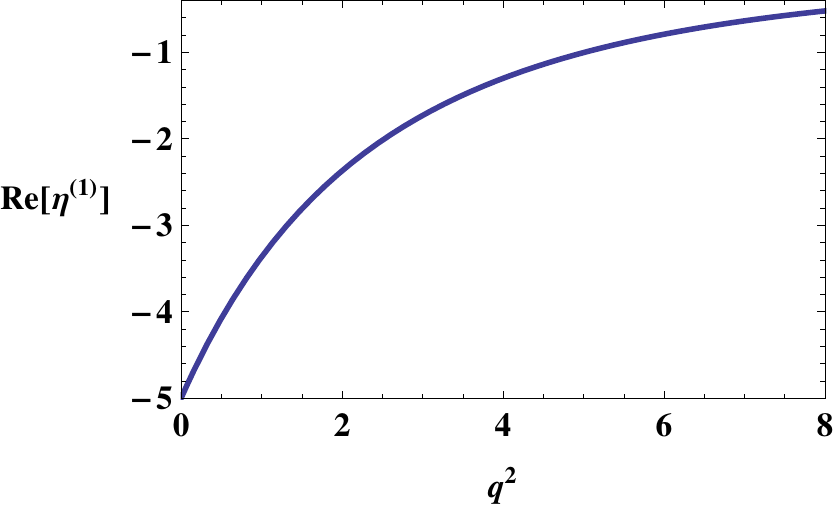}
\includegraphics[scale=0.83]{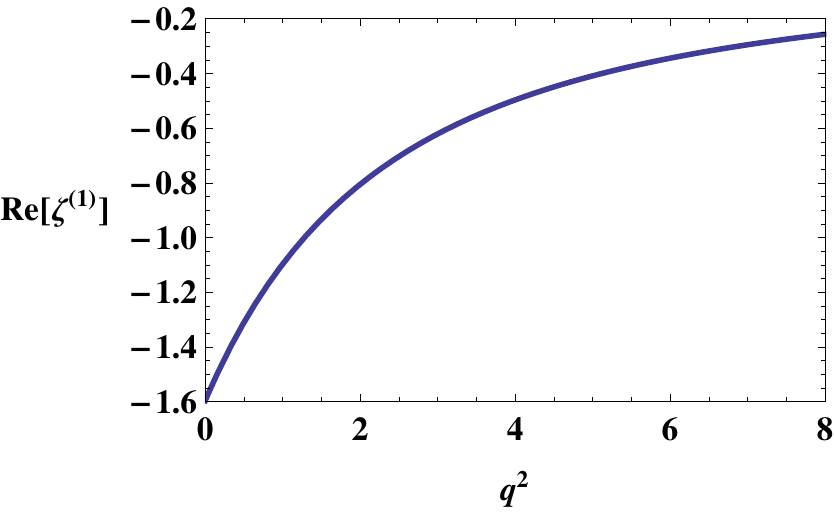}
\caption{The viscosities $\eta$ and $\zeta$ as functions of $q^2$ with $\omega=0$.}
\label{figure5}
\end{figure}
\begin{figure}[htbp]
\centering
\includegraphics[scale=0.55]{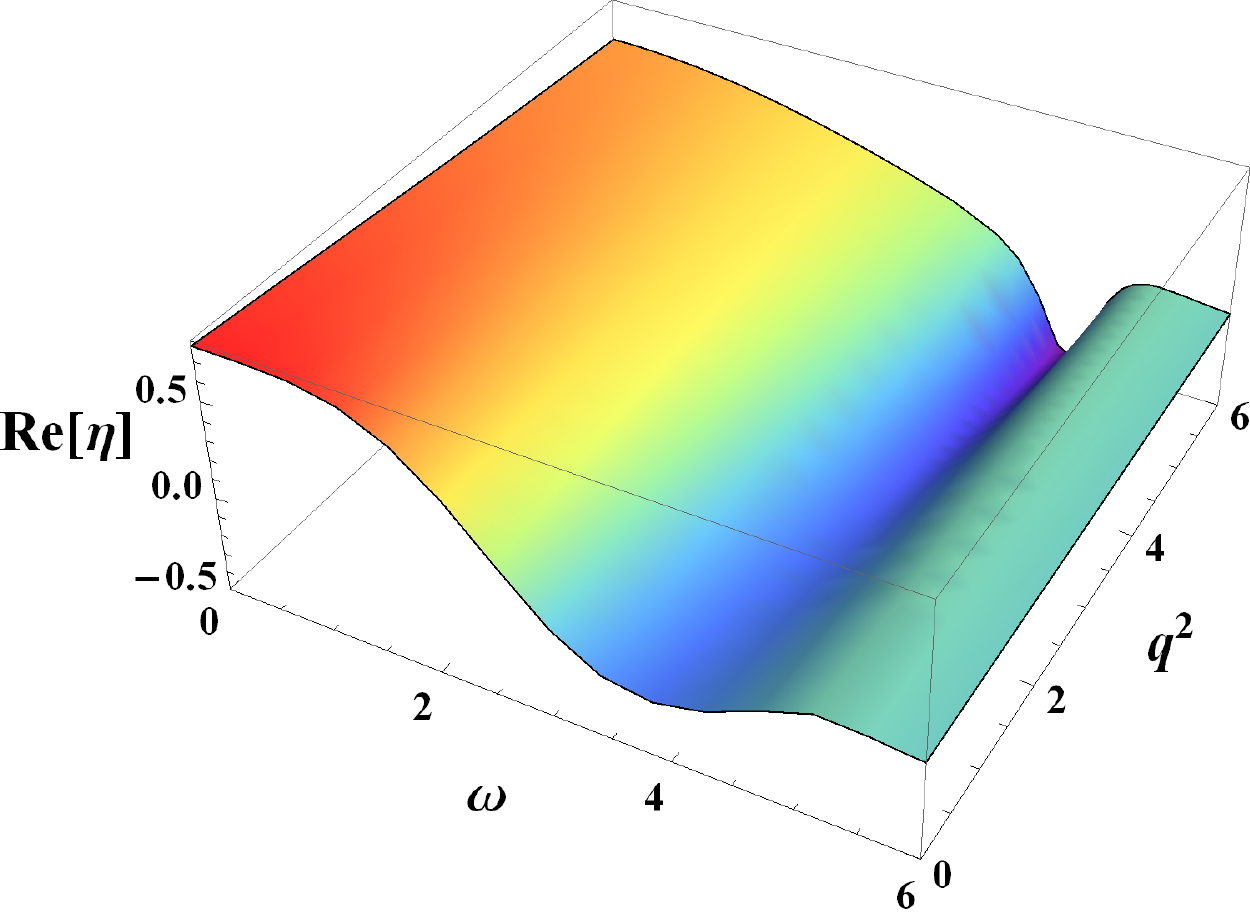}
\includegraphics[scale=0.55]{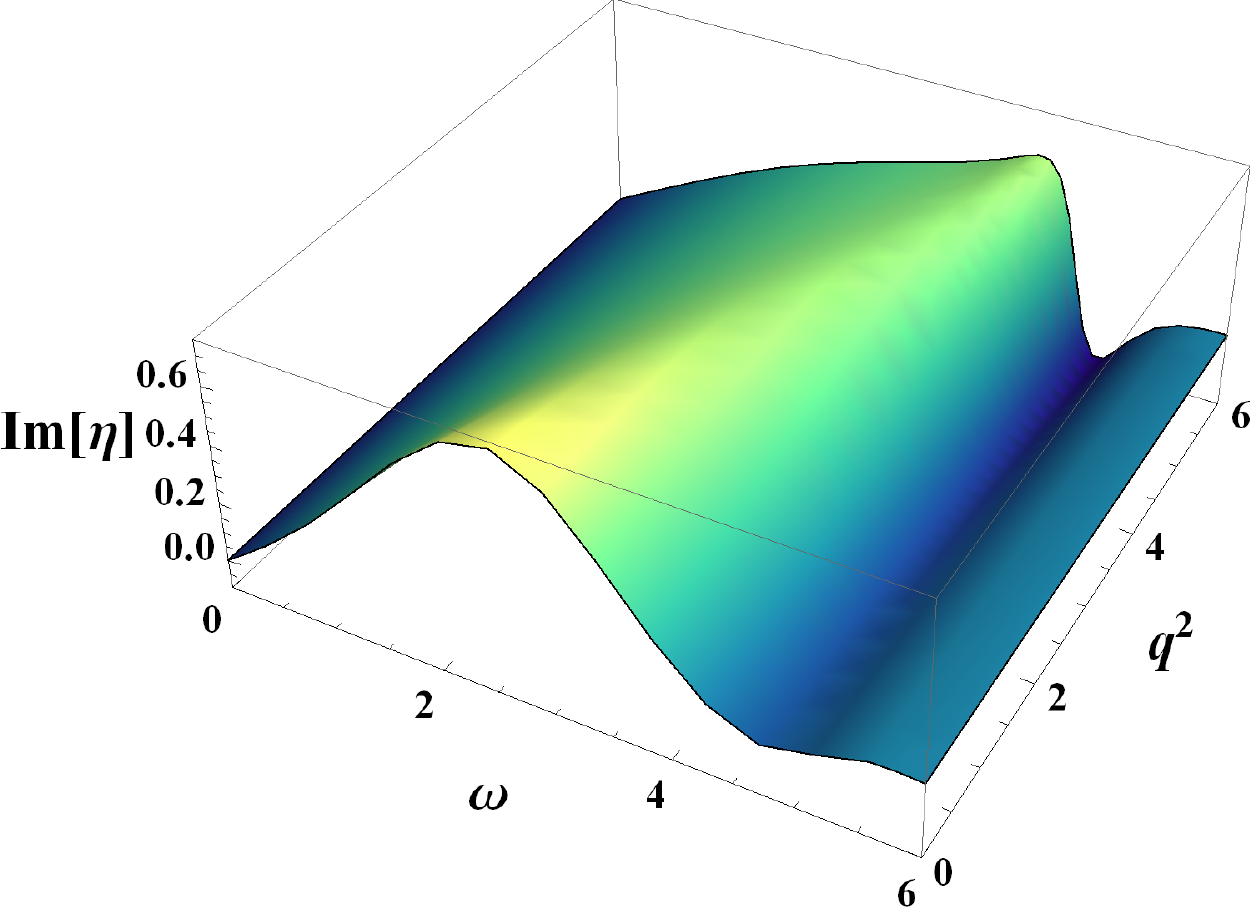}
\includegraphics[scale=0.55]{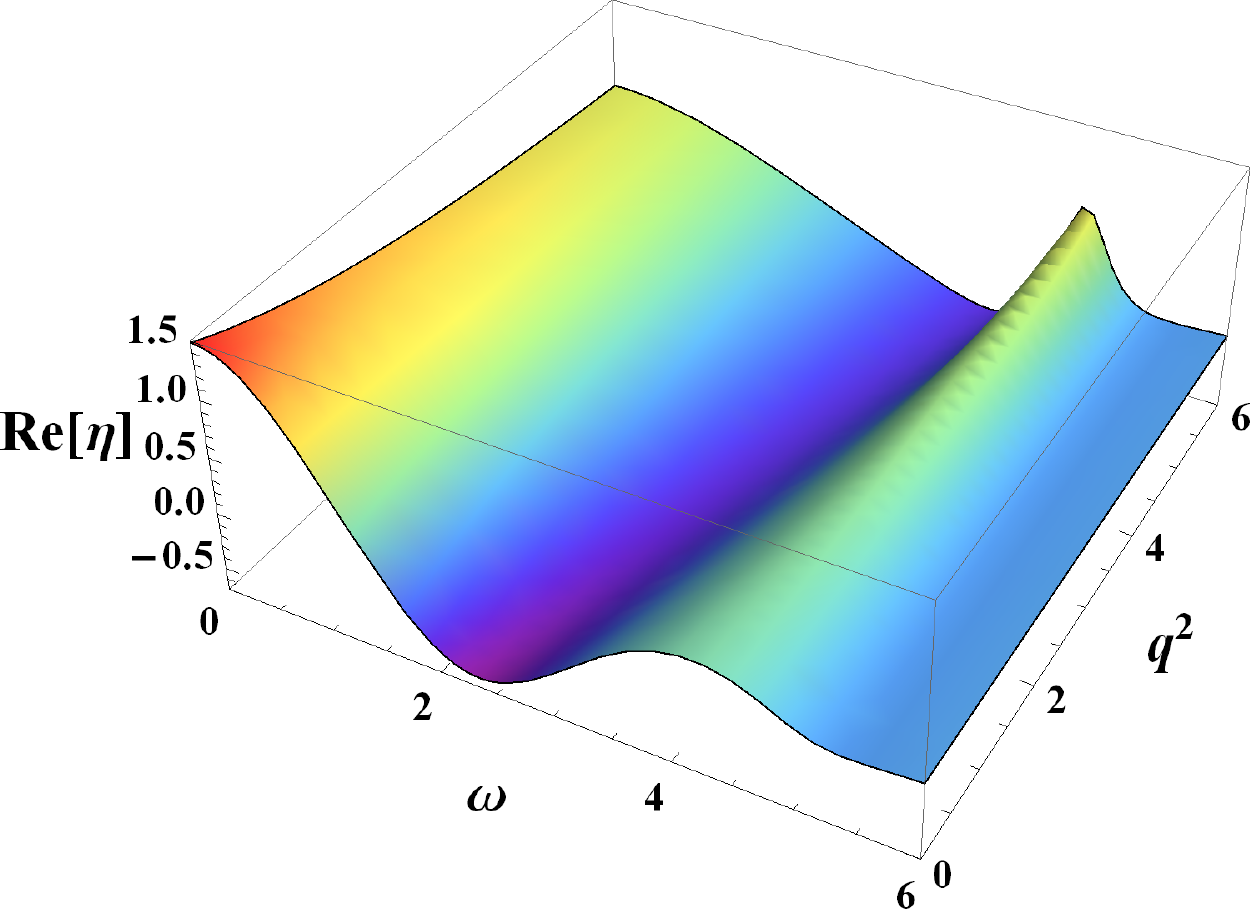}
\includegraphics[scale=0.55]{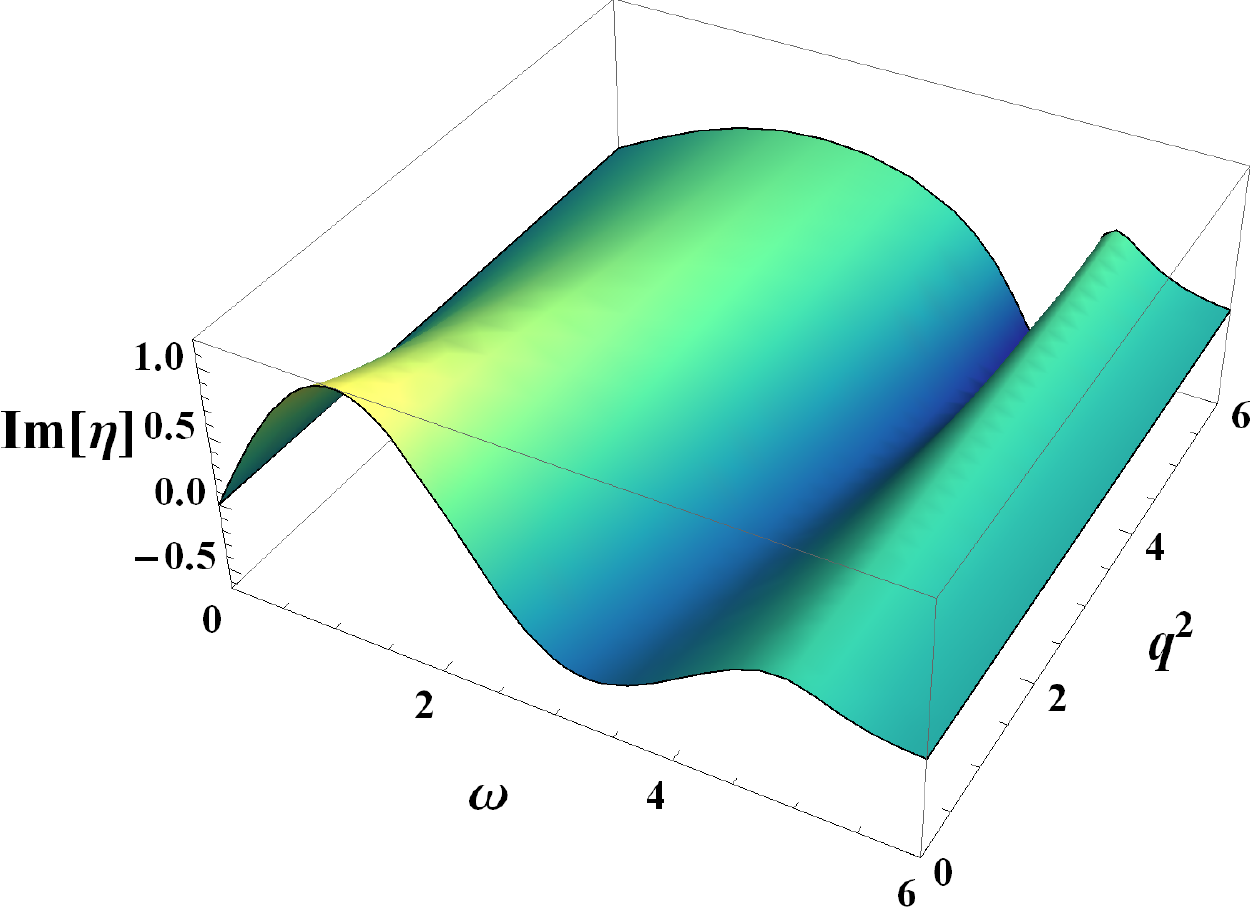}
\caption{The viscosity $\eta$ as function of $\omega$ and $q^2$ for $\alpha=9/200$ (top) and $\alpha=-7/72$ (down).}
\label{figurea1}
\end{figure}
\begin{figure}[htbp]
\centering
\includegraphics[scale=0.55]{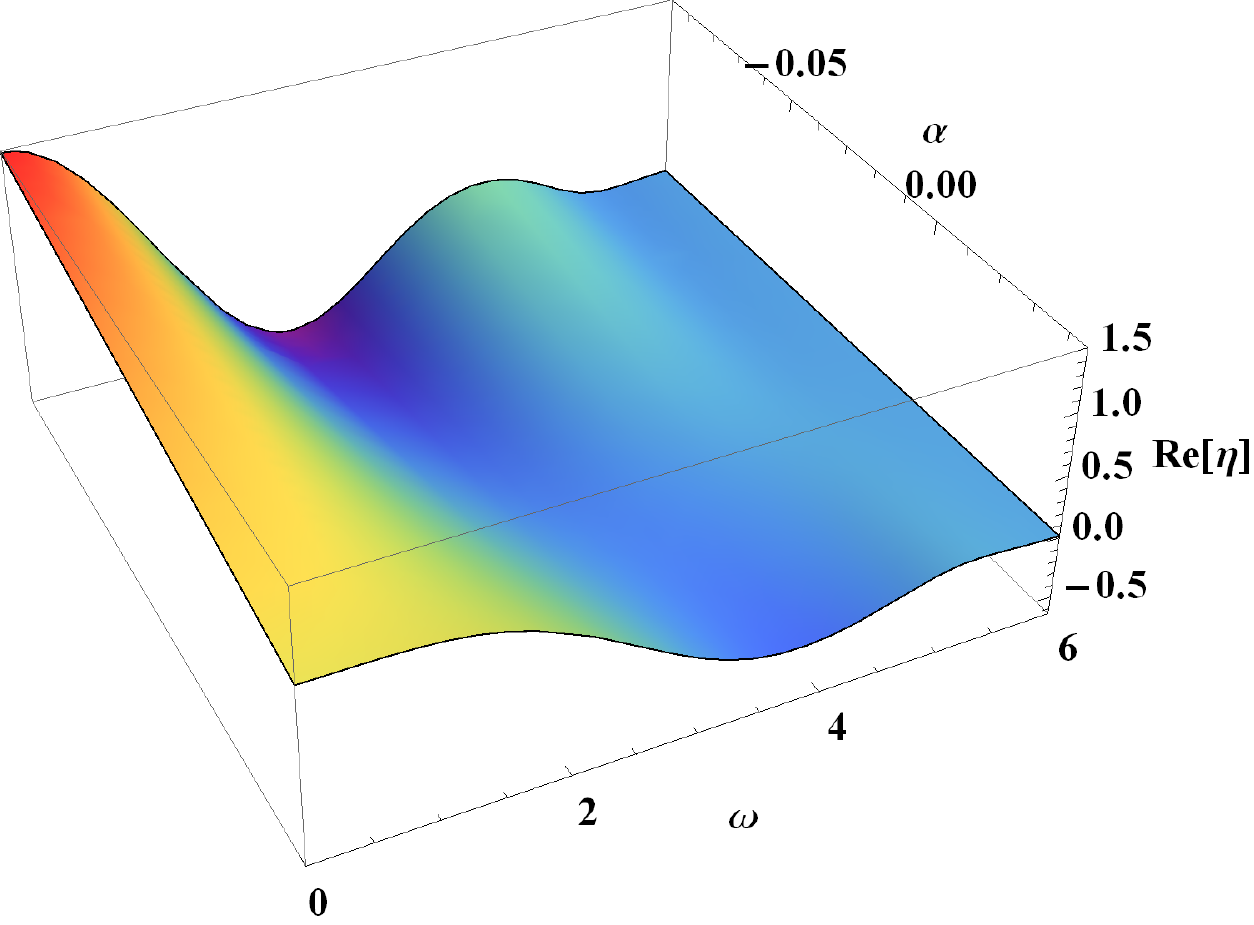}~~~
\includegraphics[scale=0.55]{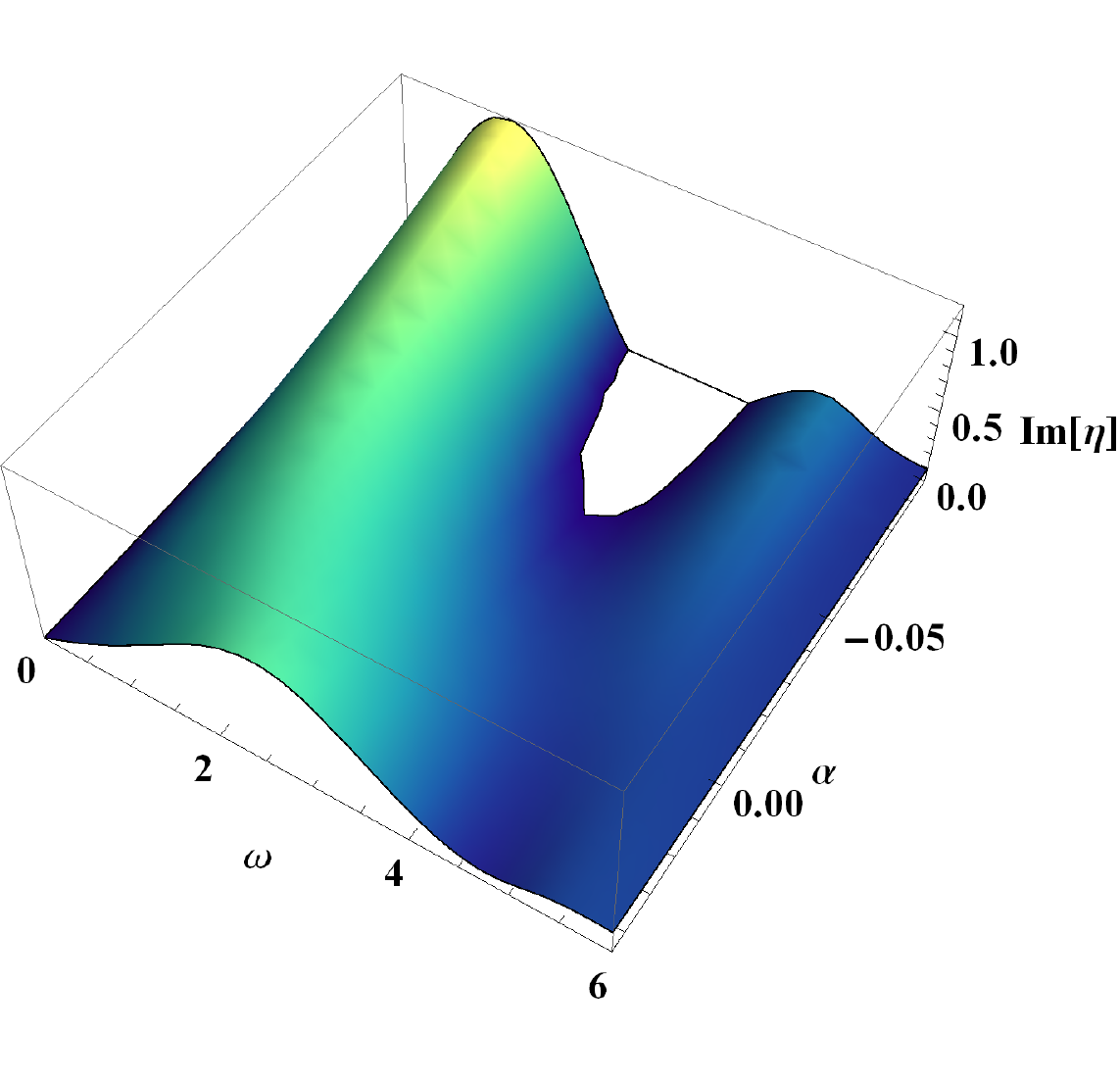}
\caption{The viscosity $\eta$ as function of $\omega$ ($q=0$) and $\alpha$ within the causality interval~(\ref{alpha bounds}). The hole around $\omega=4$ indicates the region where $\textrm{Im}[\eta]<0$.}
\label{figurea2}
\end{figure}

Numerical results for the viscosities are shown as 3D plots in Figures~\ref{figure1} and~\ref{figure2}, and then sliced at $q=0$ or $\omega=0$ in Figures~\ref{figure3},~\ref{figure4} and~\ref{figure5}. A marking behavior of all the functions is that they vanish at very large momenta, a behavior necessary for restoration of causality. Damped oscillations are clearly visible reflecting a complex pole structure of the viscosities as functions of complex $\omega$. These are the quasi-normal modes of the so-called scalar (or tensor) channel~\cite{hep-th/0506184,0905.2975,1102.4014}.

Without the Gauss-Bonnet corrections, the viscosities $\eta^{(0)}$ and $\zeta^{(0)}$ display only a weak dependence on spatial momentum $q$, meaning the dissipation is quasi-local in space. In contrast, $\eta^{(1)}$ and $\zeta^{(1)}$ introduce a much more noticeable space dependence. In order to see relative correction to viscosity function due to Gauss-Bonnet term, in Figure~\ref{figurea1} we combined $\eta^{(0)}$ and $\eta^{(1)}$ for upper and lower bounds of $\alpha$. We observed that the Gauss-Bonnet correction does introduce a profound spatial dependence for the viscosity. In addition, the Gauss-Bonnet correction changes shape of the viscosity noticeably in the intermediate regime of momenta where the amplitude of oscillation is enhanced.

Another interesting observation concerns imaginary parts of $\eta$. While $\textrm{Im}[\eta^{(0)}]$ is always positive, $\textrm{Im}[\eta^{(1)}]$ changes sign. This implies that for certain values of $\alpha$, both positive and negative, $\textrm{Im}[\eta]$ may become negative. To clearly see this behavior, in Figure~\ref{figurea2} we plot $\textrm{Im}[\eta]$ as function of $\omega$ ($q=0$) and $\alpha$ (within the causality interval~(\ref{alpha bounds})). $\textrm{Im}[\eta]$ becomes negative when $\alpha$ goes below the critical value $-0.05$. With $q$ increased, this critical value gets larger. If the viscosity function had an interpretation of a correlation function, then its imaginary part would be a spectral function and would have to be positive. Yet, beyond the first order in the gradient expansion the correlation functions get additional contributions from so-called gravitational susceptibilities of the fluid \cite{0712.2451,0905.4069,1502.08044}. So, while the possibility that $\textrm{Im}[\eta]$ becomes negative for some values of $\alpha$ does not immediately imply a problem, we take it as a signal for possible issues with causality in the theory.

To better explore the effect  of the Gauss-Bonnet corrections,  we now represent our results as memory functions in real time.
Let perform an inverse Fourier transform of $\eta(\omega,q^2)$ with respect to $\omega$ only,
\begin{equation}
\tilde{\eta}(t,q^2)=\int_{-\infty}^{\infty}\frac{d\omega}{\sqrt{2\pi}}\eta(\omega,q^2) e^{-i\omega t}.
\end{equation}
\begin{figure}[htbp]
\centering
\includegraphics[scale=0.50]{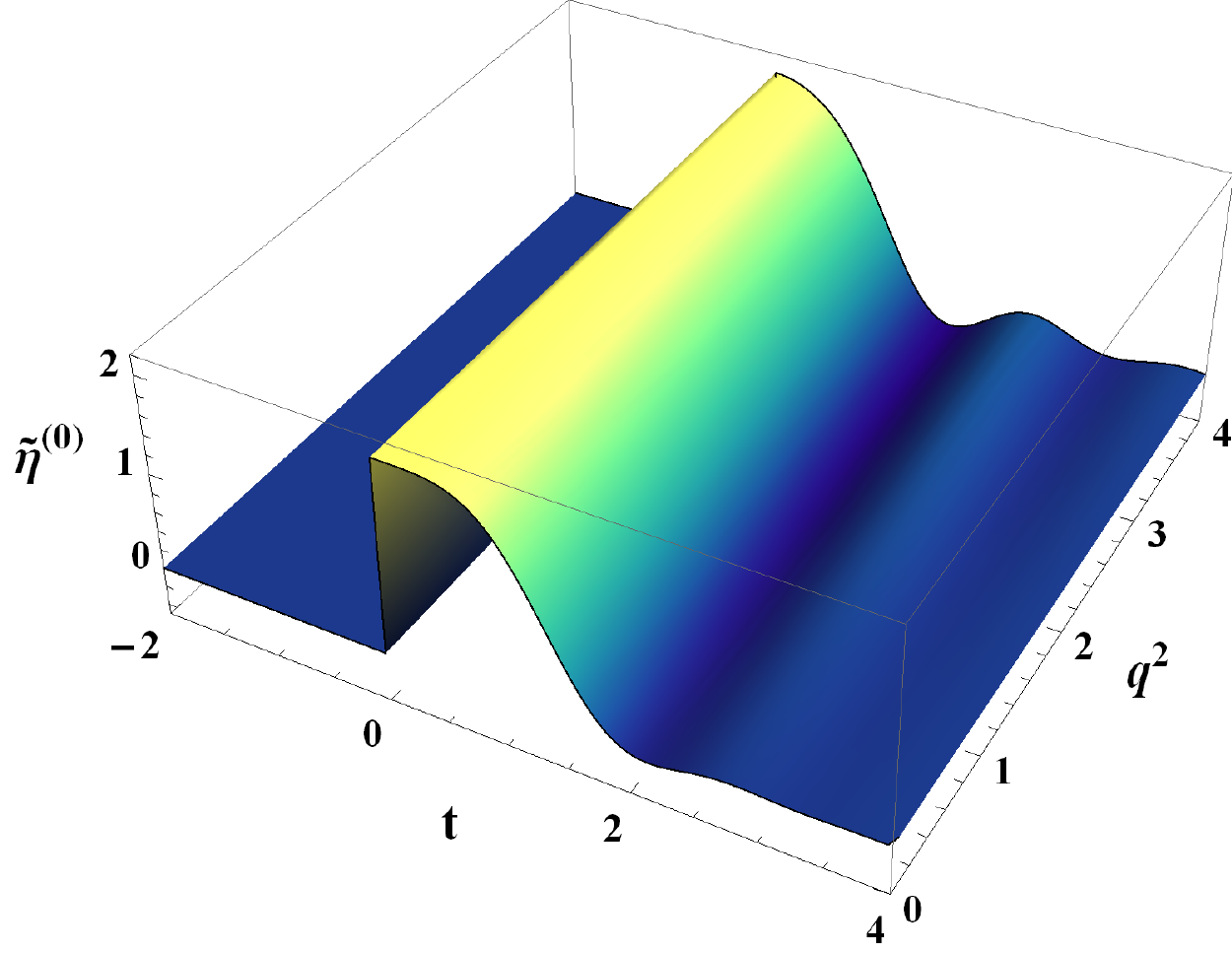}
\includegraphics[scale=0.80]{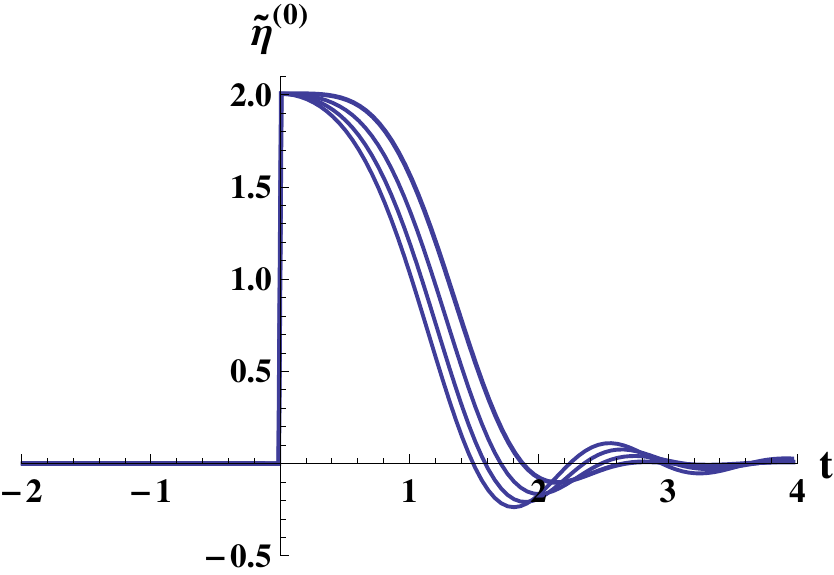}
\includegraphics[scale=0.50]{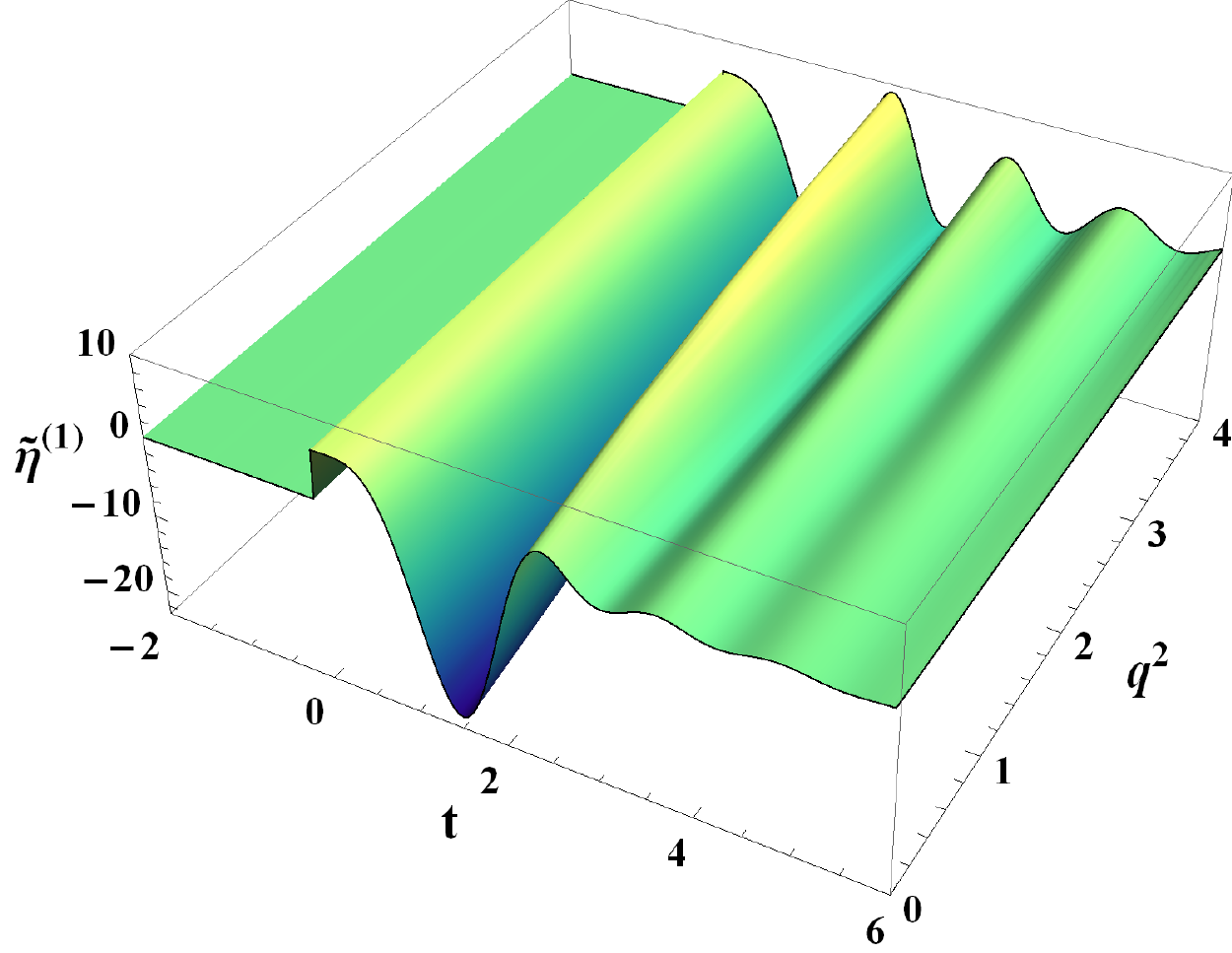}
\includegraphics[scale=0.80]{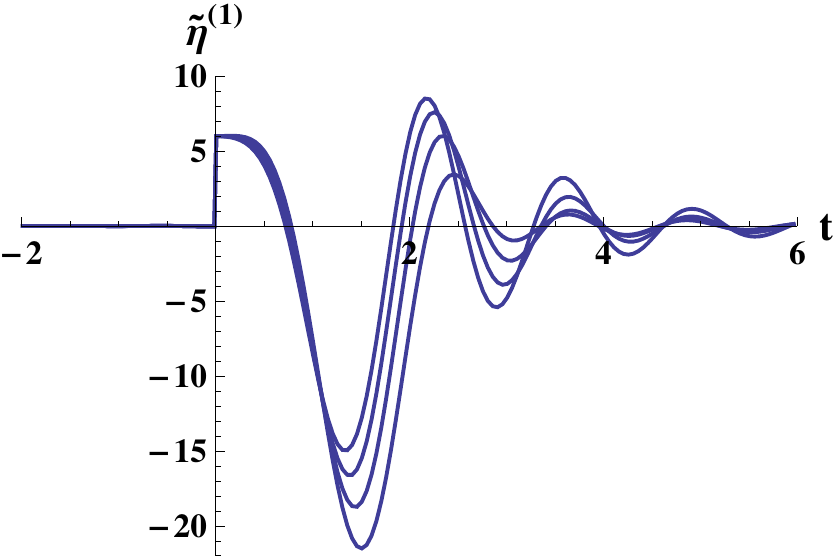}
\caption{Memory function $\tilde{\eta}(t,q^2)$. Left: 3D plot as a function of time $t$ and momentum squared $q^2$. Right: 2D plots as function of time $t$: different curves correspond to different $q^2$ ($q^2=0,1,2,3$ from the rightmost for $\tilde{\eta}^{(0)}$ and from the bottommost for $\tilde{\eta}^{(1)}$).}
\label{figure6}
\end{figure}

In Figure~\ref{figure6}, we plot the time dependence of the memory function $\tilde{\eta}(t,q^2)$. As has been pointed out in Introduction, $\tilde{\eta}^{(0)}$ vanishes for negative times, consistency with causality requirement. The Gauss-Bonnet correction $\tilde{\eta}^{(1)}$ also has support in positive times only.
 Similar effect is also found for the second memory function $\tilde{\zeta}$.
\section{Summary and discussion}\label{section4}
In this work, we discussed effects of Gauss-Bonnet corrections on holographically dual fluid dynamics. For bulk EGB gravity, we found a boosted black brane solution with locally perturbed horizon. Our construction is accurate to linear order in amplitudes of fluid velocity $u_{\mu}(x)$, temperature $T(x)$, and the Gauss-Bonnet coupling $\alpha$.
This black brane solution is dual to all-order linearly resummed fluid dynamics, and  $\alpha$-corrected viscosity functions were read off from it.

In the hydrodynamic limit, we reproduced known results for $\alpha$-corrected shear viscosity and relaxation time. As a new result, we expanded the knowledge on transport coefficients by computing $\alpha$ corrections to the third order coefficients. Beyond the hydrodynamic limit, we computed Gauss-Bonnet-corrected viscosity functions. We observe two qualitatively new effects induced by the corrections. First, the viscosities become less local in real space.  Second, due to $\alpha$ corrections, $\textrm{Im}[\eta]$ can become negative.

Finally, we Fourier transformed the viscosity functions into real time, where they play a role of memory functions. For positive times, we observed a pattern of damped oscillation reflecting a structure of complex poles. Interestingly, the poles of $\eta^{(1)}$ are apparently shifted compared to the ones of $\eta^{(0)}$. Given that the EGB memory function does not display any causality violation, we do not expect any dramatic $\alpha$-induced effects on EGB quasi-normal modes. However, we think a study of EGB quasi-normal modes might provide additional insight on the problem. This is, however, beyond the scope of the current paper.

\appendix
\section{Computational details}\label{appendix}
In this Appendix, we provide some computational details which were omitted in deriving the boundary fluid dynamics.

In terms of the metric corrections~(\ref{metric corr}), the tensor $\tilde{T}_{\mu\nu}$ is
\begin{equation}\label{stress tld1}
\begin{split}
\tilde{T}_{00}=&3(1-5\alpha)(1-4\epsilon b_{1})+\epsilon \left(3k-2r^3 \partial u +\frac{2}{r} \partial j-9r^4 h-3r^5\partial_r h- r^2\partial^2 h \right.\\
&\left.-3r^3\partial_vh +\frac{1}{2}r^2\partial_{i}\partial_{j} \alpha_{ij}\right) +\epsilon \alpha\left(8r^3\partial u-9k -\frac{12}{r}\partial j -27r^4 h + 9r^5 \partial_rh\right. \\
&\left.+5r^2\partial^2h+12r^3\partial_v h-\frac{5}{2}r^2 \partial_i\partial_j \alpha_{ij}\right),
\end{split}
\end{equation}
\begin{equation}\label{stress tld2}
\begin{split}
\tilde{T}_{i0}=&-4\epsilon (1-5\alpha)u_i +\epsilon\left(4j_i-r^3\partial_v u_i+ \frac{1}{r} \partial_i k-r\partial_r j_i-\frac{1}{2r^2}\partial^2 j_i+\frac{1}{2r^2} \partial_i\partial j\right.\\
&\left.+\frac{1}{2}r^2\partial_v\partial_k \alpha_{ik}-r^2\partial_v\partial_i h- \frac{3}{2} r^3 \partial_i h\right)+\epsilon\alpha \left(20j_i-4r^3 \partial_v u_i -\frac{7}{2r^2}\partial^2 j_i \right.\\
&\left.+ \frac{4}{r}\partial_i k -5r\partial_r j_i+\frac{7}{2r^2}\partial_i\partial j +\frac{5}{2} r^2 \partial_v\partial_k\alpha_{ik} -5r^2\partial_v\partial_i h -6r^3 \partial_ih\right),
\end{split}
\end{equation}
\begin{equation}\label{stress tld3}
\begin{split}
\tilde{T}_{ij}=&\delta_{ij}\left(1-5\alpha\right)\left(1-4\epsilon {\bf{b}}_1\right) + \epsilon \delta_{ij}\left(2r^3\partial u+9r^4h+2r^5\partial_rh-r^3\partial_v h +\frac{1}{2} r^2 \partial^2h \right.\\
&\left.-r^2\partial_v^2h+k-r\partial_r k+\frac{1}{r} \partial_v k -\frac{1}{2r^2} \partial^2k -\frac{2}{r}\partial j+\frac{1}{r^2} \partial_v \partial j -\frac{1}{2} r^2\partial_k \partial_l \alpha_{kl} \right)\\
&+\epsilon\left[\frac{1}{2r^2}\partial_i\partial_j k-r^3\left(\partial_i u_j +\partial_j u_i\right)-\frac{1}{2}r^2\partial_i \partial_j h -r^5\partial_r\alpha_{ij} -\frac{1}{2} r^2\partial^2 \alpha_{ij}\right.\\
&\left.-r^3\partial_v\alpha_{ij}+\left(\frac{1}{r}-\frac{1}{2r^2}\partial_v\right) \left(\partial_i j_j +\partial_j j_i\right)+\frac{1}{2}r^2\left(\partial_i \partial_k \alpha_{jk} +\partial_j\partial_k \alpha_{ik}\right) \right.\\
&\left.+\frac{1}{2}r^2\partial_v^2 \alpha_{ij} \right]-\epsilon\alpha \delta_{ij} \left(8r^3\partial u+3k-3r\partial_r k+\frac{4}{r} \partial_v k -\frac{5}{2r^2} \partial^2k-27r^4h\right.\\
&\left.+ 3r^5 \partial_rh +4r^3\partial_v h+\frac{5}{2}r^2\partial^2h -5r^2\partial_v^2h +\frac{7}{r^2}\partial_v\partial j -\frac{12}{r}\partial j -\frac{5}{2}r^2 \partial_k \partial_l\alpha_{kl}\right)\\
&-\epsilon \alpha \left[\frac{5}{2r^2}\partial_i\partial_j k
-4r^3\left(\partial_i u_j+\partial_j u_i \right)-\frac{5}{2}r^2\partial_i \partial_j h +\left(\frac{6}{r}-\frac{7}{2r^2}\partial_v\right)\left(\partial_i j_j+\partial_j j_i \right) \right.\\
&\left.-3r^5\partial_r\alpha_{ij}-4r^3\partial_v\alpha_{ij}-\frac{5}{2}r^2\partial^2 \alpha_{ij} +\frac{5}{2} r^2 \partial_v^2\alpha_{ij}+\frac{5}{2}r^2 \left(\partial_i\partial_k\alpha_{jk}+\partial_j \partial_k\alpha_{ik}\right) \right],
\end{split}
\end{equation}
where we have dropped  terms that explicitly vanish at $r=\infty$.

For consistency, constraints in~(\ref{field eq}) have to be satisfied by the gravity solution presented in section~\ref{section3}. We find it more convenient to consider suitable combinations of $E_{MN}=0$. The first one is $E_{vv}+r^2f(r)E_{vr}=0$,
\begin{equation}\label{constraint1}
\begin{split}
0=&\;4r^3\partial u+r^2\partial^2{\bf{b}}_1-12r^3\partial_v{\bf{b}}_1- \left(r^2+ r^6\right) \partial_r\partial u- 4r^3\partial j-r^2 \partial^2k +3r^3 \partial_vk \\
&+ 2r^2\partial_v \partial j +\left(r^4-1\right)\partial_r\partial j-\frac{\alpha}{r^2} \left[2r^5\left(3r^4+7\right)\partial u+24r^4 \partial^2 {\bf{b}}_1-60r^5\partial_v {\bf{b}}_1\right.\\
&\left.-\left(3r^8+3r^4-2\right)\partial_v\partial u-2\left(11r^5-r\right)\partial j- 4 \left(r^4-1\right)\partial^2k- \left(r^8-r^4\right)\partial_i\partial_j\alpha_{ij} \right.\\
&\left.+3\left(3r^5-4r\right)\partial_vk+4\left(3r^4-1\right)\partial_v\partial j- 3 \left(r^7-r^3\right)\partial_rk + 4\left(r^6-2r^2\right)\partial_r\partial j\right].
\end{split}
\end{equation}
The combination $E_{vi}+r^2f(r)E_{ri}=0$ yields
\begin{eqnarray}\label{constraint2}
\begin{split}
0=&\;r^4\partial^2 u_i-r^4\partial_i\partial u+4r\partial_i{\bf{b}}_1- 4r\partial_v u_i- r^4\partial_v^2 u_i-\partial^2 j_i +\partial_i\partial j- r\partial_i k+4r \partial_v j_i \\
&+r^4\partial_v\partial_k\alpha_{ik}+\left(r^6-r^2\right)\partial_r\partial_k\alpha_{ik}
+r^2\partial_r\partial_i k -r^2\partial_r\partial_v j_i-\frac{\alpha}{r^4} \left[\left( 3r^9-23r^5\right)\partial_v u_i \right.\\
&\left.+20r^5 \partial_i {\bf{b}}_1+\left(5r^8+3r^4 \right)\left(\partial^2 u_i - \partial_i \partial u \right)-4\left(r^8-r^4\right)\partial_v^2 u_i - \left(3r^5-20r\right)\partial_i k\right.\\
&\left.- 6\left(r^4+1\right) \left(\partial^2j_i-\partial_i \partial j\right)+20r^5\partial_v j_i-3 \left(r^7-r^3\right)\partial_r j_i+\left(3r^6-4r^2\right) \partial_r\partial_i k\right.\\
&\left.- \left(5r^6-2r^2\right)\partial_r\partial_v j_i+\left(r^8-r^4\right)\partial_r^2 j_i+2 \left(2r^{10}-r^6-3r^2\right) \partial_r\partial_k \alpha_{ik}\right.\\
&\left.+4\left(r^8+r^4\right) \partial_v\partial_k\alpha_{ik}\right].
\end{split}
\end{eqnarray}
With the near $r=\infty$ behaviors~(\ref{asym abcd},\ref{asym k}) at hand, the large $r$ limit of~(\ref{constraint1},\ref{constraint2}) can be shown to produce the conservation law $\partial^{\mu}T_{\mu\nu}=0$.

In the hydrodynamic limit, we perturbatively solved holographic RG flow equations~(\ref{abcd momentum}). Recall the formal expansion~(\ref{abcd expansion})
\begin{equation}
\begin{split}
a\left(\bar{\omega},\bar{q}_i,r\right)&=\sum_{n=0}^{\infty}\lambda^n a_n\left(\bar{\omega},\bar{q}_i,r\right),~~~ b\left(\bar{\omega},\bar{q}_i,r\right)=\sum_{n=0}^{\infty}\lambda^n b_n\left(\bar{\omega},\bar{q}_i,r\right),\\
c\left(\bar{\omega},\bar{q}_i,r\right)&=\sum_{n=0}^{\infty}\lambda^n c_n\left(\bar{\omega},\bar{q}_i,r\right),~~~ d\left(\bar{\omega},\bar{q}_i,r\right)=\sum_{n=0}^{\infty}\lambda^n d_n\left(\bar{\omega},\bar{q}_i,r\right).
\end{split}
\end{equation}
Then, perturbative solutions for the metric corrections can be expressed as double integrals. We here summarize the main results,
\begin{equation}
a_0=0,
\end{equation}
\begin{equation}
\begin{split}
c_0=&-\int_{r}^{\infty}\frac{dx}{\left(x^5-x\right)-2\alpha \left(x^5-3x^{-3} \right)} \int_{r_{H}}^x\left[-3y^2+\alpha\left(9y^2-4y^{-2}\right)\right]dy\\
&\xlongrightarrow{r\to \infty}\frac{1-\alpha}{r}-\frac{1-8\alpha}{4r^4}+\mathcal{O} \left(\frac{1}{r^5}\right),
\end{split}
\end{equation}
\begin{equation}
a_1=-i\bar{\omega}\left[\left(1+\alpha\right)r^3+\frac{2\alpha}{r}\right],
\end{equation}
\begin{equation}
\begin{split}
c_1=&\;-\int_{r}^{\infty}\frac{dx}{x^5-x-2\alpha\left(x^5-3x^{-3}\right)} \int_{r_{H}}^x dy \left\{3i\bar{\omega} y^2c_0(y)+2i\bar{\omega} y^3\partial_y c_0(y)-i\bar{\omega} y\right.\\
&~~~~~~~~~~\left.-\alpha\, i\bar{\omega} \left[\left(9y^2-4y^{-2}\right)c_0(y)+ 2 \left(3y^3+4y^{-1}\right) \partial_y c_0(y) -4y+12y^{-3}\right]\right\}\\
&\;\xlongrightarrow{r\to\infty}-\frac{i\bar{\omega}}{8r^4}\left[2-\ln{2}-\alpha\left(23-6\ln{2} \right)\right]+\mathcal{O}\left(\frac{1}{r^5}\right),
\end{split}
\end{equation}
\begin{equation}
\begin{split}
b_0=&\;-\int_{r_{H}}^rx^3\,dx\int_x^{\infty} dy \frac{y-y^3\partial_y c_0(y)/3+\alpha\left[4\left(y^3+y^{-1}\right)\partial_y c_0(y) -\left(5y+y^{-3}\right)\right]}{y^4-\alpha\left(5y^4-2\right)}\\
&\;-\frac{3}{8}+\frac{2}{3}\alpha\xlongrightarrow{r\to\infty}-\frac{1}{3}r^2+ \mathcal{O}\left(\frac{1}{r}\right),
\end{split}
\end{equation}
\begin{eqnarray}
\begin{split}
a_2=&\;\int_r^{\infty}dx\,x^3\int_x^{\infty}dy\frac{\bar{q}^2y^3\partial_y c_0(y) + \bar{q}^2 y - \alpha\left[4\bar{q}^2\left(y^3+y^{-1}\right)\partial_y c_0(y) +\bar{q}^2\left(5y+y^{-3}\right) \right]}{y^4-\alpha\left(5y^4-2\right)}\\
&\xlongrightarrow{r\to\infty} \frac{1}{5r}\bar{q}^2\left(1-7\alpha\right)+\mathcal{O} \left(\frac{1}{r^2}\right),
\end{split}
\end{eqnarray}
\begin{equation}
\begin{split}
d_0=&\;-\int_r^{\infty}\frac{dx}{x^5-x-2\alpha\left(x^5-3x^{-3}\right)} \int_{r_H}^x \left\{\frac{2}{y}\partial_y b_0(y)-\frac{2}{y^2}b_0(y)+\frac{2}{3}yc_0(y)\right.\\
&~~~~~~~~~~\left.-\alpha \left[\left(\frac{10}{y}+\frac{12}{y^5}\right)\partial_y b_0 (y) -\left(\frac{10}{y^2}+\frac{60}{y^6}\right)b_0(y) + \frac{8}{3}\left(y-3y^{-3}\right) c_0(y) \right]\right\}\\
&\;\xlongrightarrow{r\to\infty} -\frac{1}{48r^4}\left[5-\pi-2\ln{2}-\alpha \left(82-14\pi-28\ln{2}\right)\right]+\mathcal{O}\left(\frac{1}{r^5}\right),
\end{split}
\end{equation}
\begin{eqnarray}
\begin{split}
c_2=&-\int_r^{\infty}\frac{dx}{x^5-x-2\alpha\left(x^5-3x^{-3}\right)}\int_{r_H}^x dy \left\{2i\bar{\omega} y^3\partial_y c_1(y)+3i\bar{\omega} y^2 c_1(y) + y^{-1}\partial_y a_2(y) \right.\\
&~~~~~~~~~\left.-y^{-2}a_2(y)-\alpha\left[2i\bar{\omega} \left(3y^3+4y^{-1}\right) \partial_y c_1(y) + i\bar{\omega} \left(9y^2-4y^{-2}\right)c_1(y)\right.\right.\\
&~~~~~~~~~\left.\left. -5\left(y^{-2}+6y^{-6}\right)a_2(y) + \left(5y^{-1}+6y^{-5}\right) \partial_y a_2(y)\right]\right\},
\end{split}
\end{eqnarray}
where $r_H=1-\alpha$ as defined in~(\ref{horizon}). From  large $r$ behavior of these functions, we arrive at the power expansion~(\ref{pert exp}) of the viscosity functions.

\section*{Acknowledgements}
We would like to thank I. Arefeva, G. Beuf, R. Brustein, A. Buchel, R. Janik, A. Kovner, J. Maldacena, G. Policastro,  K. Skenderis and A. Starinets for informative discussions related to this work. We thank the Galileo Galilei Institute for Theoretical Physics for the hospitality and the INFN for financial support during the workshop ``Holographic Methods for Strongly Coupled Systems'' where this work was completed. This work was supported by the ISRAELI SCIENCE FOUNDATION grant \#87277111, BSF grant \#012124, the People Program (Marie Curie Actions) of the European Union's Seventh Framework under REA grant agreement \#318921; and the Council for Higher Education of Israel under the PBC Program of Fellowships for Outstanding Post-doctoral Researchers from China and India (2014-2015).

\providecommand{\href}[2]{#2}\begingroup\raggedright\endgroup

\end{document}